\documentclass{iopconfser}
\usepackage{graphicx}
\usepackage{amssymb,amsmath,cite}
\usepackage[percent]{overpic}  
\usepackage{xcolor,xspace}
\usepackage{braket}
\usepackage{wasysym}  

\newcommand{\ba}{\begin{eqnarray}}
\newcommand{\ea}{\end{eqnarray}}
\newcommand{\ban}{\begin{eqnarray*}}
\newcommand{\ean}{\end{eqnarray*}}
\newcommand{\bsub}{\begin{subequations}}
\newcommand{\esub}{\end{subequations}}

\def\ket#1{|#1\rangle}

\def\b0{\beta_0}
\def\beq{\beta_{\rm eq}}
\def\g0{\gamma_0}
\def\gaeq{\gamma_{\rm eq}}

\def\ack{\section*{Acknowledgments}}

\begin{document}

\title{Intertwined Quantum Phase Transitions in Bose and
  Bose-Fermi Systems}

\author{A. Leviatan}

\affil{Racah Institute of Physics, The Hebrew University,
  Jerusalem 91904, Israel}

\email{ami@phys.huji.ac.il}

\begin{abstract}
Pronounced
structural changes within individual configurations
(Type~I QPT), superimposed on an abrupt crossing of
these configurations (Type~II QPT), define the
notion of intertwined quantum phase transitions (QPTs).
We discuss and present evidence for
such a scenario
in finite Bose and Bose-Fermi
systems. The analysis is based on
algebraic models with explicit configuration mixing,
where the two types of QPTs describe 
shape-phase transitions in-between
different dynamical symmetries and shape-coexistence
with crossing.
\end{abstract}

\section{Introduction}

\hspace{12pt}
Quantum phase transitions (QPTs) are
qualititative changes in the
properties of a physical system induced by variation of
parameters in the
Hamiltonian~\cite{Gilmore1978, Gilmore1979}.
They occur at zero temperature hence are driven by
quantum fluctuations and
their study is currently a topic
of great interest in diverse fields~\cite{Carr2010QPT}.
Such ground-state phase transitions can be categorized
into two types.
The first, denoted Type~I QPT, is a phase transition in
a single configuration~\cite{CejJolCas10},
described by an Hamiltonian of
the form~\cite{diep80},
\ba
\label{eq:type-i}
\hat H(\xi) = (1-\xi)\hat H_1 + \xi\,\hat H_2 ~,
\ea
where the separate Hamiltonians
$\hat{H}_1$ and $\hat{H}_2$ are non-commuting,
act in the same Hilbert space (${\cal H}$)
and correspond to different phases of the system.
When the control parameter $\xi$ changes from $0$ to $1$,
the spectra and eigenstates of $\hat{H}(\xi)$ progress
from those of $\hat{H}_1$ to those of $\hat{H}_2$.
The number of basis states is preserved throughout the
transition. A schematic illustration of the evolving
spectra in a Type~I QPT is shown in Fig.~1(a).

\hspace{12pt}
A second type of phase transition, denoted Type~II QPT, 
occurs when two (or more) configurations
coexist~\cite{Heyde11}.
In this case,
the quantum Hamiltonian can be cast
in matrix form~\cite{Frank2006},
\ba
\label{eq:type-ii}
\hat{H}(\xi_A,\xi_B,\omega) &=&
\left [
\begin{array}{cc}
\hat{H}_{A}(\xi_A) & \hat{W}(\omega) \\ 
\hat{W}(\omega) & \hat{H}_{B}(\xi_B)
\end{array}
\right ] ~,\quad
\label{Hmat}
\ea
where the indices $A$ and $B$ denote the two
configurations. The terms,
\bsub
\ba
\hat{H}_A(\xi_A) &=&
(1-\xi_A)\hat{H}_{A;1} + \xi_A\,\hat{H}_{A;2} ~,\\
\hat{H}_B(\xi_B) &=&
(1-\xi_B)\hat{H}_{B;1} + \xi_B\,\hat{H}_{B;2} ~,
\ea
\esub
act in different Hilbert spaces
(${\cal H}_A$ and ${\cal H}_B$),
and the mixing term
$\hat{W}(\omega)$ connects the two spaces.
As the control parameters $(\xi_A,\xi_B,\omega)$
are varied, each of the sub-Hamiltonians,
$\hat{H}_A(\xi_A)$ and $\hat{H}_B(\xi_B)$,
can undergo a QPT of Type~I, and the combined Hamiltonian
$\hat{H}(\xi_A,\xi_B,\omega)$, which acts in the
enlarged Hilbert space,
${\cal H} = {\cal H}_A\otimes{\cal H}_B$,
can experience a Type~II QPT with
a crossing of the two configurations.
In this case, states with distinct properties
coexist at similar energies,
as shown schematically in Fig.~1(b).
Usually, the individual Type~I QPTs are masked by
the strong mixing between the states in
the separate configurations.
However, if the mixing is weak, the abrupt Type~II QPT
is accompanied by a distinguished Type~I QPT within
each configuration separately. Such a scenario, referred
to as intertwined QPTs~\cite{Gav2019,Gav2020,Gav2022},
and its manifestation in
Bose and Bose-Fermi systems is the subject matter of
the present contribution.
\begin{figure}[t!]
\centering
\hspace{-0.7cm}
\begin{overpic}[width=\linewidth]{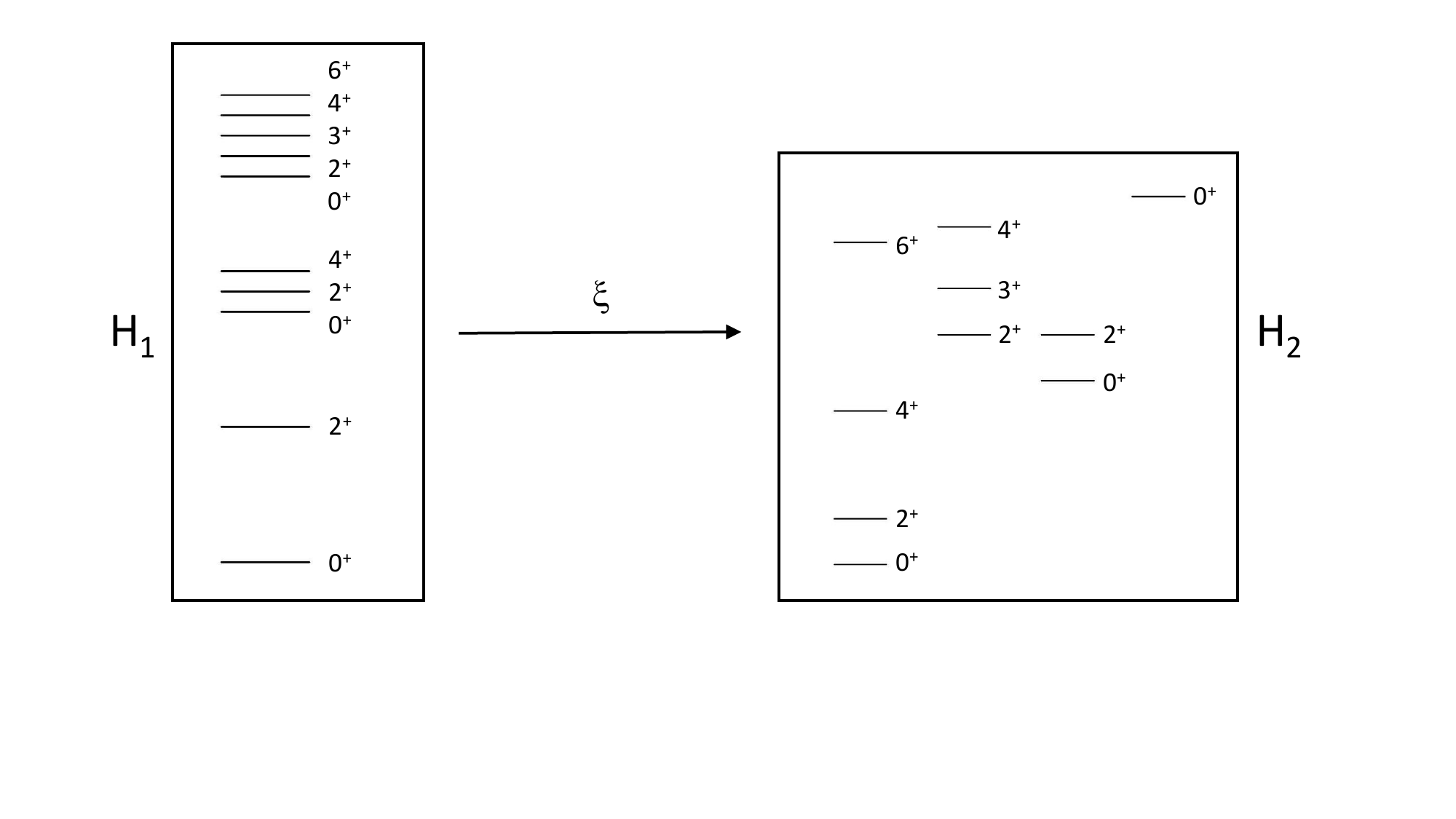}
\put (31,48) {\large(a) Type I QPT}
\end{overpic}\\
\vspace{-1.8cm}
\hspace{-0.4cm}
\begin{overpic}[width=\linewidth]{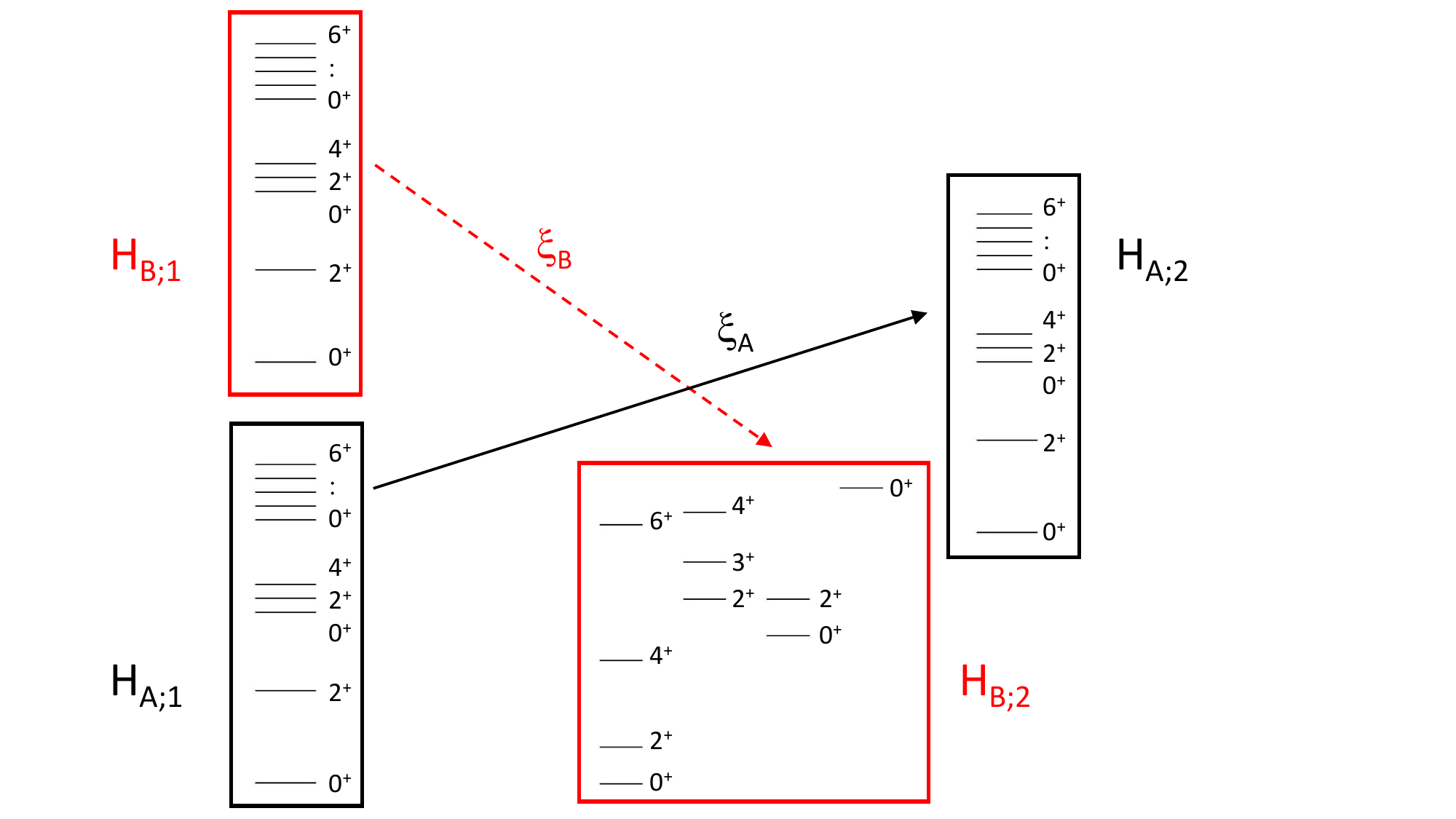}
\put (31,50) {\large(b) Type II QPT}
\end{overpic}
\caption{\label{fig1-QPTs}
\small
Schematic illustration for the evolution of structure
in a quantum phase transition (QPT).
(a)~Type~I QPT, Eq.~(\ref{eq:type-i}),
portraying a change from
spherical-vibrator to deformed-rotor spectra within
the Hilbert space of a single configuration.
(b)~Type~II QPT, Eq.~(\ref{eq:type-ii}),
portraying a coexistence and crossing of
sets of levels associated with
different configurations ($A$ and $B$).
When both types of QPTs are distinguishable,
they define
intertwined QPTs in the extended Hilbert space.}
\end{figure}

\section{Algebraic approach to quantum phase transitions
in Bose systems}

\hspace{12pt}
Algebraic models based on compact spectrum generating
algebras are particularly useful for studying QPTs in
mesoscopic (finite) systems. They are endowed with a rich
symmetry structure, amenable for both quantum and
classical analysis and provide a simple framework for
tractable yet detailed calculation of observables.
As a representative for this class of models
in bosonic systems, 
we consider an interacting boson model for a single
configuration and its extensions to multiple
configurations.

\subsection{The interacting boson model}

\hspace{12pt}
The interacting boson model (IBM)~\cite{ibm},
describes low-lying
quadrupole collective states in
even-even nuclei in terms of $N$ interacting
monopole $(s)$ and quadrupole $(d)$ bosons.
The model is based on a U(6) spectrum generating algebra
with elements
${\cal G}_{ij}\equiv b^{\dag}_{i}b_j =
\{s^{\dag}s,\,s^{\dag}d_{m},\,
d^{\dag}_{m}s,\,d^{\dag}_{m}d_{m '}\}$.
The IBM Hamiltonian is expanded in terms of these
generators, 
$\hat{H} = \sum_{ij}\epsilon_{ij}\,{\cal G}_{ij} 
+ \sum_{ijk\ell}u_{ijk\ell}\,{\cal G}_{ij}{\cal G}_{k\ell}$, 
and consists of Hermitian, 
rotational-invariant interactions 
which conserve the total number of $s$- and $d$- bosons, 
$\hat N = \hat{n}_s + \hat{n}_d = 
s^{\dagger}s + \sum_{m}d^{\dagger}_{m}d_{m}$. 
A dynamical symmetry (DS) occurs if the Hamiltonian
can be written in terms of the Casimir operators 
of a chain of nested sub-algebras of U(6).
The Hamiltonian is then completely solvable in the basis
associated with each chain. 
The solvable limits of the IBM correspond to the
following DS chains,
\bsub
\ba
&&{\rm U(6)\supset U(5)\supset SO(5)\supset SO(3)}
\;\qquad
\quad\ket{N,\, n_d,\,\tau,\,n_{\Delta},\,L} ~,
\label{U5-ds}
\\
&&{\rm U(6)\supset SU(3)\supset SO(3)} \;\;\;
\qquad\quad\qquad
\;\;\;\ket{N,\, (\lambda,\mu),\,K,\, L} ~,
\label{SU3-ds}
\\
&&
{\rm U(6)\supset SO(6)\supset SO(5)\supset SO(3)} \qquad
\;\;\,\ket{N,\, \sigma,\,\tau,\,n_{\Delta},\, L} ~.
\label{SO6-ds}
\ea
\label{IBMchains}
\esub
The basis members indicated above, are 
classified by the irreducible representations (irreps)
of the corresponding algebras.
Specifically, the quantum numbers
$N,n_d,(\lambda,\mu),\sigma,\tau,L$,
label the relevant irreps of
U(6), U(5), SU(3), SO(6), SO(5), SO(3), 
respectively, and $n_{\Delta},K$ are multiplicity labels.
The basis states are eigenstates of the DS Hamiltonian
with energies expressed by the eigenvalues of the
Casimir operators in a given chain.

\hspace{12pt}

In the U(5)-DS limit, Eq.~(\ref{U5-ds}),
the spectrum resembles that of a spherical vibrator,
where states are arranged in U(5) $n_d$-multiplets at
approximately equal energy spacing. The ground and
lowest excited states are
$(n_d\!=\!0,\, L\!=\!0)$ and
$(n_d=\!1\!,\, L\!=\!2)$, 
$(n_d\!=\!2,\, L\!=\!0,2,4)$,
$(n_d\!=\!3, L\!=\! 2,0,3,4,6)$.
The spectra in the SU(3)-DS limit, Eq.~(\ref{SU3-ds}),
resembles that of an axially-deformed prolate rotor
with states arranged in SU(3) $(\lambda,\mu)$-multiplets
forming rotational bands with $L(L+1)$-splitting. 
The lowest irrep $(2N,0)$ contains the ground band
$g(K\!=\!0)$.
The first excited irrep $(2N\!-\!4,2)$ contains 
both the $\beta(K\!=\!0)$
and $\gamma(K\!=\!2)$ bands.
The lowest members of each $K$-band have
$L=0,2,4\ldots$ ($L=2,3,4\ldots$) for
$K\!=\!0$ ($K\!=\!2$).
The spectrum in the
SO(6)-DS limit of Eq.~(\ref{SO6-ds}),
resembles that of a $\gamma$-unstable
deformed rotor, where states are arranged 
in SO(6) $\sigma$-multiplets forming rotational bands
with $\tau(\tau+3)$ splitting.
The lowest irrep $\sigma\!=\!N$ contains the
ground ($g$) band and the first excited irrep
$\sigma\!=\!N-2$ contains the $\beta$-band.
The lowest members in each band have quantum numbers 
$(\tau\!=\!0,\, L\!=\!0)$, $(\tau\!=\!1,\, L\!=\!2)$, 
$(\tau\!=\!2,\, L\!=\!2,4)$ and
$(\tau\!=\!3,\, L\!=\!0,3,4,6)$.
Characteristic energy ratios
for the U(5), SU(3), SO(6) limits are
$R_{4/2}= E(4_1)/E(2_1)=2,\,3.33,\,2.5$,
respectively.

\hspace{12pt}
Geometry is introduced in the algebraic model by means
of a coset space 
$U(6)/U(5)\otimes U(1)$  and
an energy surface, 
\ba
E_{N}(\beta,\gamma) &=&
\langle \beta,\gamma; N\vert \hat{H}
\vert \beta,\gamma ; N\rangle ~,
\label{enesurf}
\ea
obtained by the expectation value of the Hamiltonian
in a `projective' coherent state~\cite{gino80,diep80},
\bsub
\ba
\vert\beta,\gamma ; N \rangle &=&
(N!)^{-1/2}(b^{\dagger}_{c})^N\,\vert 0\,\rangle ~,\\
b^{\dagger}_{c} &=& (1+\beta^2)^{-1/2}[\beta\cos\gamma 
d^{\dagger}_{0} + \beta\sin{\gamma} 
( d^{\dagger}_{2} + d^{\dagger}_{-2})/\sqrt{2} + s^{\dagger}] ~.
\ea
\label{int-state}
\esub
Here $(\beta,\gamma)$ are
quadrupole shape parameters whose values,
$(\beta_{\rm eq},\gamma_{\rm eq})$, 
at the global minimum of $E_{N}(\beta,\gamma)$
define the equilibrium 
shape for a given Hamiltonian.
The equilibrium deformations associated with the 
DS limits, Eq.~(\ref{IBMchains}),
conform with their geometric interpretation,
\begin{subequations}
\begin{align}
  \text{U(5)}: & \quad
  \beq = 0 \qquad\qquad\qquad\qquad
  \text{spherical shape}~,\\
  \text{SU(3)}: &\quad
  (\beta_{\rm eq} = \sqrt{2},\gamma_{\rm eq}\!=\!0)
  \qquad\quad
 \text{prolate-deformed shape}~,\\  
 \text{SO(6)}: & \quad
 (\beta_{\rm eq} = 1,\,\gaeq\,\text{arbitrary})
 \quad
 \text{$\gamma$-unstable deformed shape}~.
\end{align}
\label{DS-shapes}
\end{subequations}
The coherent state
$\ket{\beq,\gaeq;N}$ of Eq.~(\ref{int-state}),
with the equilibrium deformations,
serves as an intrinsic state for the ground band,
whose rotational members are obtained by angular
momentum projection.

\subsection{Microscopic consideration}

\hspace{12pt}
A microscopic interpretation of the IBM is based on its
link with the nuclear shell model~\cite{IacTal87}.
According to it,
the $s$ and $d$ bosons represent valence nucleon pairs
with total number
$N\!=\!N_{\pi}+N_{\nu}$, where $N_{\pi}$ ($N_{\nu}$)
is the number of proton (neutron) particle or hole pairs
counted from the nearest closed shell.
The states constructed of valence nucleons are referred
to as normal states. The Type~I QPTs discussed
in Section~2.3.
are structural changes involving a subset of
normal states represented by a single configuration
of $N$ bosons. In nuclei near closed
shells, additional configurations associated with
particle-hole excitations across shell gaps,
play a role. The resulting states are referred
to as intruder states~\cite{Heyde11}.
The latter can drop in energy and coexist with normal
states due to correlations induced by the residual
interactions~\cite{FedPit79,HeyCas85}.
Type~II QPTs involving the crossing
of normal and intruder states, can be modeled in the IBM
by incorporating additional configurations with different
boson numbers, in the manner described in Section~2.4.

\subsection{Type~I QPT}

\hspace{12pt}
In the algebraic approach to QPTs, the dynamical symmetries
correspond to possible shape-phases of the system
in accord with their geometric interpretation.
QPTs in a single configuration (Type~I) can be studied
by an Hamiltonian $\hat{H}(\xi)$,
as in Eq.~(\ref{eq:type-i}), which
interpolates between the different DS limits (phases)
by varying the control parameter
$\xi$~\cite{diep80,iaczam04}.
The energy surface, $E_{N}(\xi;\beta,\gamma)$ of
Eq.~(\ref{enesurf}), depends parametrically
on $\xi$ and
serves as the classical mean-field Landau potential.
The nature of the phase transition and critical points
are determined by the derivatives with respect to $\xi$ of
the energy surface, evaluated at the equilibrium
deformations.
The order of the QPT is the order of the derivative
where discontinuities first occur
(Ehrenfest classification)~\cite{Gilmore1979}.

\hspace{12pt}
A typical Hamiltonian frequently used for Type I QPTs
has the form,
\begin{equation}\label{eq:ham-q}
\hat H(\epsilon_d,\kappa,\chi) =
\epsilon_d\, \hat n_d
+ \kappa\, \hat Q_\chi \cdot \hat Q_\chi~,
\end{equation}
where the quadrupole operator is given by,
\begin{equation}\label{eq:q-op}
\hat Q_\chi =
d^\dag s+s^\dag \tilde d \!+\!\chi
(d^\dag \times \tilde d)^{(2)} ~.
\end{equation}
Here $\tilde d_m = (-1)^m d_{-m}$ and standard
notation of angular momentum coupling is used.
The control parameters $(\epsilon_d,\kappa,\chi)$
in Eq.~\eqref{eq:ham-q} with values $(\kappa\!=\!0)$,
$(\epsilon_d\!=\!0,\chi\!=\!-\sqrt{7}/2)$ and
$(\epsilon_d\!=\!0,\chi\!=\!0)$, interpolate between
the respective U(5), SU(3) and SO(6) DS~limits.
For the Hamiltonian~\eqref{eq:ham-q}, the
associated Landau potential~(\ref{enesurf}) reads,
\ba
\label{eq:surface-single}
E_{N}(\beta,\gamma;\epsilon_d,\kappa,\chi) =
5\kappa\, N + \frac{N\beta^2}{1+\beta^2} 
\left[\epsilon_d + \kappa (\chi^2-4)\right]
+ \frac{N(N-1)\beta^2}{(1+\beta^2)^2}\kappa
\left[4 - 4\bar{\chi}\beta\,\cos3\gamma
  + \bar\chi^2\beta^2\right] ~,
\ea
where $\bar\chi\!=\!\sqrt{\frac{2}{7}}\chi$.
The U(5)-SU(3) transition is found to be first-order,
the U(5)-SO(6) transition is second order and the
SU(3)-SO(6) transition is a crossover.
The order parameter is taken to be the
expectation value of the $d$-boson number operator,
$\hat n_d$, in the ground state.
The latter converges in the large-$N$ limit
to the classical order parameter related to the
equilibrium deformation $\beta_{\rm eq}$,
\begin{equation}\label{eq:order-param}
  \frac{\braket{\hat n_d}_{0^+_1}}{N}
\approx \frac{\beta_{\rm eq}^2}{1+\beta_{\rm eq}^2}~.
\end{equation}
IBM Hamiltonians of the above form have been used
extensively for studying shape-phase transitions, 
exemplifying Type~I QPTs in
nuclei~\cite{CejJolCas10,iaczam04,jolie09,iac11,Macek14}.

\subsection{Type~II QPT}

\hspace{12pt}
The effect of additional configurations, representing
intruder states in nuclei, can be studied in the
framework of the interacting boson model with configuration
mixing (IBM-CM)~\cite{Duval1981,Duval1982}. The latter is
an extension of the IBM 
based on associating the different shell-model
spaces of 0p-0h, 2p-2h, 4p-4h,$\dots$ particle-hole
excitations, with the corresponding boson spaces
comprising of
$N,\, N\!+\!2,\, N\!+\!4,\ldots$ bosons, which are
subsequently mixed.
For two configurations, the IBM-CM Hamiltonian
has the form as in Eq.~(\ref{Hmat}) with entries,
\bsub
\label{eq:ham_ab}
\ba
\hat H_A(\xi_A)
& = & \hat H(\epsilon^{A}_{d},\kappa_{A},\chi) ~,
\label{eq:ham_a}\\
\hat H_B(\xi_B)
& = & \hat H(\epsilon^{B}_{d},\kappa_{B},\chi)
+ \kappa^{\prime}_{B} \hat L \cdot \hat L + \Delta ~,
\label{eq:ham_b}\\
\hat W(\omega) & = & \omega\, [\,(d^\dag\times d^\dag)^{(0)}
  + (s^\dag)^2\,] + {\rm H.c.} ~,
\label{eq:mixing}
\ea
\esub
where H.c. stands for Hermitian conjugate.
The Hamiltonian $\hat{H}_{A}$ represents
the normal $A$ configuration ($N$ boson space)
and $\hat{H}_{B}$ represents the intruder
$B$ configuration ($N\!+\!2$ boson space).
They involve terms similar to those of the
single-configuration Hamiltonian of Eq.~(\ref{eq:ham-q}).
$\hat{H}_{B}$ contains  
an additional rotational
term and an off-set $\Delta$  between the
two configurations.
The resulting eigenstates $\ket{\Psi;L}$
are linear combinations of the wave functions,
$\Psi_A$ and $\Psi_B$, in the two spaces $[N]$ and
$[N\!+\!2]$,
\begin{equation}
  \label{eq:wf}
  \ket{\Psi; L} =
  a\ket{\Psi_A; [N], L} + b\ket{\Psi_B; [N\!+\!2], L}
\;\; , \;\; a^2+ b^2 =1 ~.
\end{equation}
The normal and intruder components of $\ket{\Psi; L}$
can be expanded in the any of the DS bases
of Eq.~(\ref{IBMchains}) with the appropriate
boson numbers,
\ba
\ket{\Psi; L} =
\sum_{\alpha} C^{(N,L)}_{\alpha}\ket{N,\alpha,L}
+\sum_{\alpha} C^{(N+2,L)}_{\alpha}\ket{N+2,\alpha,L} ~,
\label{eq:wf2}
\ea
where
$\alpha$ denotes $(n_d,\,\tau,\,n_{\Delta})$, 
$[(\lambda,\mu),K]$ and
$(\sigma,\,\tau,\,n_{\Delta})$
for the U(5), SU(3) and SO(6) chains,
respectively. From such expansions one can extract the
probabilities for a given symmetry species, 
\begin{subequations}\label{eq:decomp-ds}
\begin{align}
  \text{U(5)}: \quad & P^{(N,L)}_{n_d}
  = \sum_{\tau,n_\Delta}[C^{(N,L)}_{n_d,\tau,n_\Delta}]^2
  \;\;\;,\;\;\;
  P^{(N+2,L)}_{n_d}
  = \sum_{\tau,n_\Delta}[C^{(N+2,L)}_{n_d,\tau,n_\Delta}]^2 ~,
\label{eq:decomp-u5}\\
  \text{SU(3)}: \quad & P^{(N,L)}_{(\lambda,\mu)}
  = \sum_{K}[C^{(N,L)}_{(\lambda,\mu),K}]^2
  \;\;\;,\;\;\;
  P^{(N+2,L)}_{(\lambda,\mu)}
= \sum_{K}[C^{(N+2,L)}_{(\lambda,\mu),K}]^2 ~,
\label{eq:decomp-su3}\\
  \text{SO(6)}: \quad & P^{(N,L)}_{\sigma}
  = \sum_{\tau,n_\Delta}[C^{(N,L)}_{\sigma,\tau,n_\Delta}]^2
  \;\;\;,\;\;\;
  P^{(N+2,L)}_{\sigma}
  = \sum_{\tau,n_\Delta}[C^{(N+2,L)}_{\sigma,\tau,n_\Delta}]^2 ~.
\label{eq:decomp-so6}
\end{align}
\end{subequations}
The above DS decompositions highlight the symmetry
and shape content, Eq.~(\ref{DS-shapes}), of the states
considered. In particular,
spherical-type of states are dominated by
a single $n_d$ component in the U(5)
probabilities, Eq.~(\ref{eq:decomp-u5}),
while deformed type of states
show a large spread in $n_d$.
The probabilities of the normal-intruder mixing
for the state $\ket{\Psi;L}$,
Eqs.~(\ref{eq:wf})-(\ref{eq:wf2}),
are given by,
\ba
a^2  = \sum_{\alpha} |C^{(N,L)}_{\alpha}|^2
\;\;\; , \;\;\;
b^2  = \sum_{\alpha} |C^{(N+2,L)}_{\alpha}|^2
= 1-a^2 ~.
\label{Prob-ab}
\ea
The $E2$ operator in the IBM-CM reads
$\hat{T}(E2) = e_A\, \hat Q^{(N)}_{\chi}
+ e_B\, \hat Q^{(N+2)}_{\chi}$,
where $\hat Q_\chi$ is defined in Eq.~\eqref{eq:q-op},
a superscript ($N$) denotes a projection onto the
$[N]$ boson space and  ($e_A,\,e_B$) are boson
effective charges.

\hspace{12pt}
A geometric interpretation~\cite{Frank2004}
is obtained by means of the matrix 
$E(\beta,\gamma)$,
\begin{align}\label{eq:surface-mat}
E(\beta,\gamma) =
\left [
\begin{array}{cc}
E_A(\beta,\gamma;\xi_A) & \Omega(\beta,\gamma;\omega) \\ 
\Omega(\beta,\gamma;\omega) & E_B(\beta,\gamma;\xi_B)
\end{array}
\right ] ,
\end{align} 
whose entries are the matrix elements of the corresponding
terms in the Hamiltonian~\eqref{eq:type-ii}, between  the
intrinsic states~\eqref{int-state} of the two
configurations, with appropriate boson numbers.
The explicit expressions are found to be,
\begin{subequations}\label{eq:surface-elements}
\begin{align}
E_A(\beta,\gamma;\xi_A) & = 
E_N(\beta,\gamma;\epsilon^{A}_d,\kappa_{A},\chi)~,
\label{eq:e_a}
\\
E_B(\beta,\gamma;\xi_B)  & = 
E_{N+2}(\beta,\gamma;\epsilon^{B}_d,\kappa_{B},\chi)
+ 6\kappa^{\prime}_{B}\frac{(N+2)\beta^2}{1+\beta^2}
+ \Delta ~,
\label{eq:e_b}
\\
\Omega(\beta,\gamma;\omega) & = 
\frac{\sqrt{(N+2)(N+1)}}{1+\beta^2}
\omega\Bigg(1 + \frac{1}{\sqrt{5}}\beta^2\Bigg)~,
\label{eq:e_w}
\end{align}
\end{subequations}
where the surfaces on the right-hand-side of
Eqs.~(\ref{eq:e_a})-(\ref{eq:e_b})
are obtained from
Eq.~\eqref{eq:surface-single}.
Diagonalization of this two-by-two matrix produces
the so-called eigen-potentials, $E_{\pm}(\beta,\gamma)$.

\hspace{12pt}
$E(\beta,\gamma)$ of Eq.~(\ref{eq:surface-mat})
depends on  the Hamiltonian's parameters
$(\xi_A,\xi_B,\omega)$ and
in Type II QPTs, serves as the Landau potential
matrix~\cite{Frank2006}. 
The order parameters are taken to be the expectation
value of $\hat n_d$ in the ground state wave function,
$\ket{\Psi;L=0^+_1}$, and in its $\Psi_A$ and $\Psi_B$
components~(\ref{eq:wf}), denoted by
$\braket{\hat n_d}_{0^+_1}$,
$\braket{\hat n_d}_A$ and $\braket{\hat n_d}_B$,
respectively.
As can be inferred from Eq.~(\ref{eq:order-param}),
$\braket{\hat n_d}_A$ and $\braket{\hat n_d}_B$
provide information
on the shape of each configuration, $A$ and $B$.
$\braket{\hat n_d}_{0^+_1}$ involves
their sum weighted by the probabilities
of each component,
\begin{equation}
\label{eq:order-param-cm}
\braket{\hat n_d}_{0^+_1} =
a^2\braket{\hat n_d}_A + b^2\braket{\hat n_d}_B ~,
\end{equation}
and portrays the effect of the normal-intruder mixing
on the shape of the ground state.

\hspace{12pt}
IBM-CM Hamiltonians and surfaces of the above form
have been used extensively for studying shape-coexistence
and crossing of different configurations, exemplifying
Type~II QPTs in nuclei near (sub-) shell
closure~\cite{Duval1981, Duval1982,Frank2004, 
  Sambataro1982, Ramos2014, Nomura2016c,Gav2019,
  Gav2020,Gav2022,Ramos2022,GavLev2023}.
\begin{figure}[t]
\centering
\includegraphics[width=0.7\linewidth]{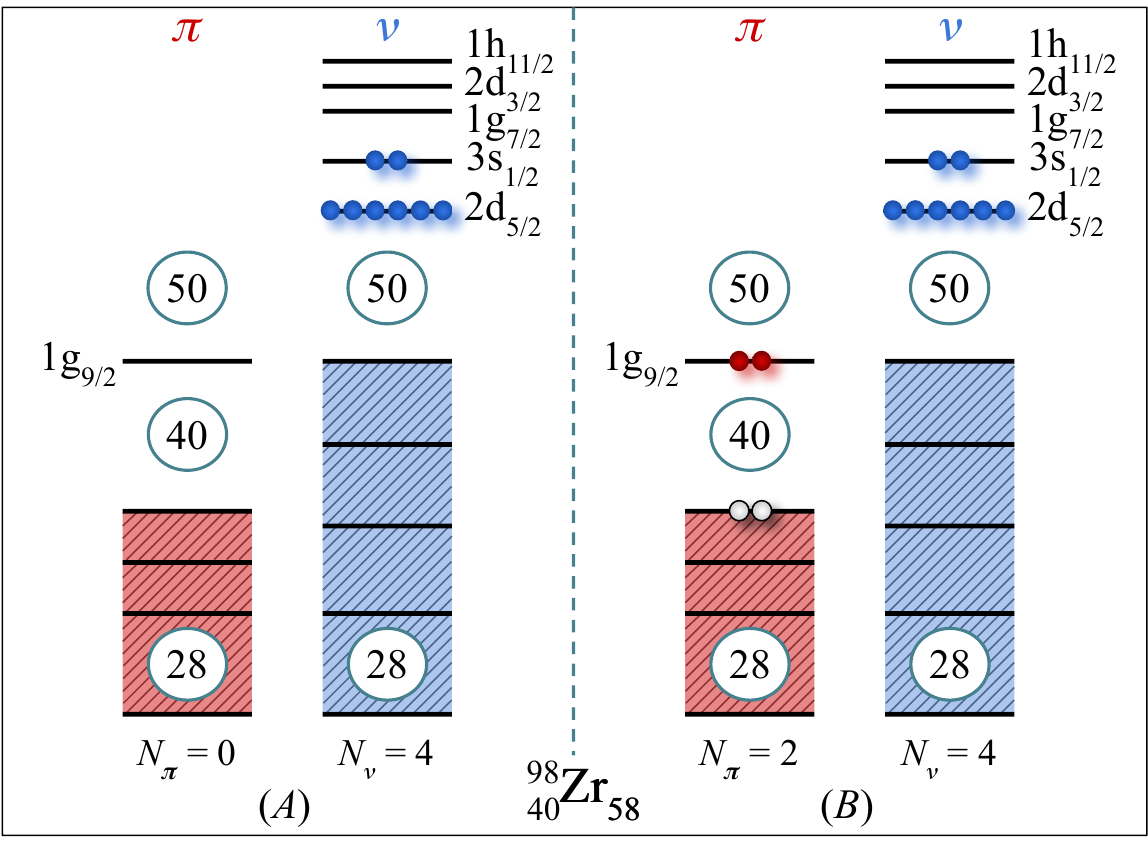}
\caption{
\small
Schematic representation of the two shell-model
configurations ($A$ and $B$) for $^{98}_{40}$Zr$_{58}$.
The corresponding numbers of proton bosons ($N_{\pi}$)
and neutron bosons ($N_{\nu}$)
are listed for each configuration and $N=N_{\pi}+N_{\nu}$.
\label{fig2:zr-98-shell}}
\end{figure}
\begin{figure*}[t]
\begin{overpic}[width=0.49\linewidth]{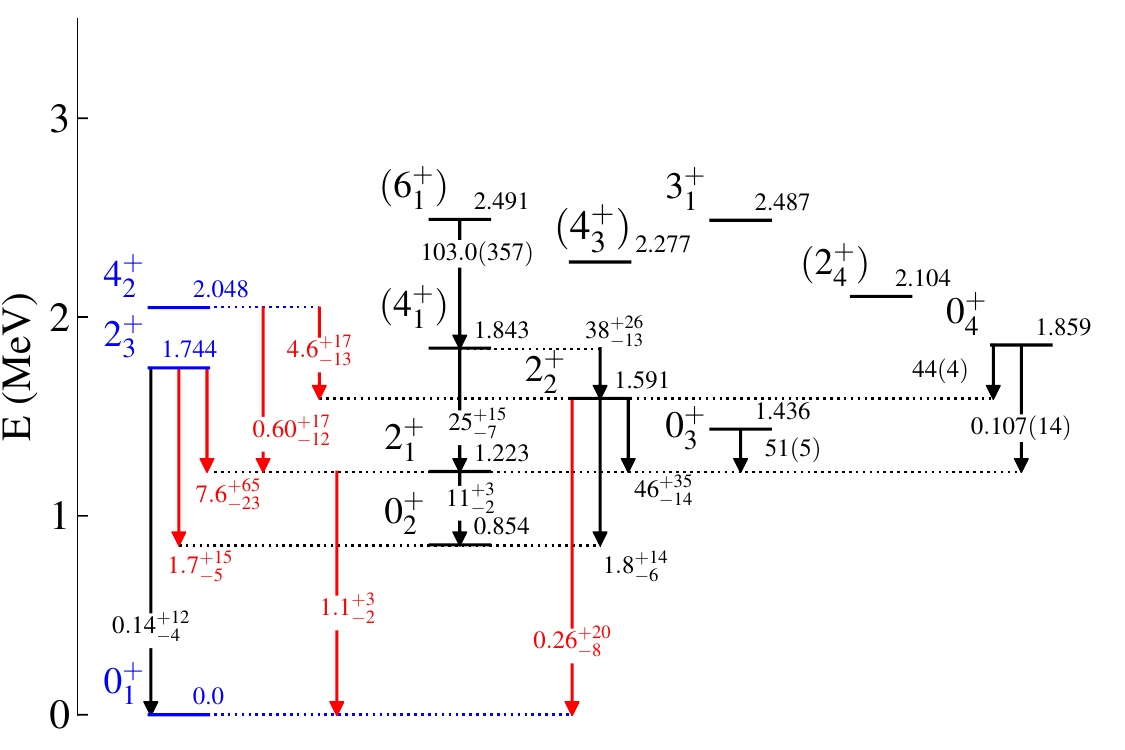}
\put (70,60) {\large (a) {\bf $^{98}$Zr exp}}
\end{overpic}
\begin{overpic}[width=0.49\linewidth]{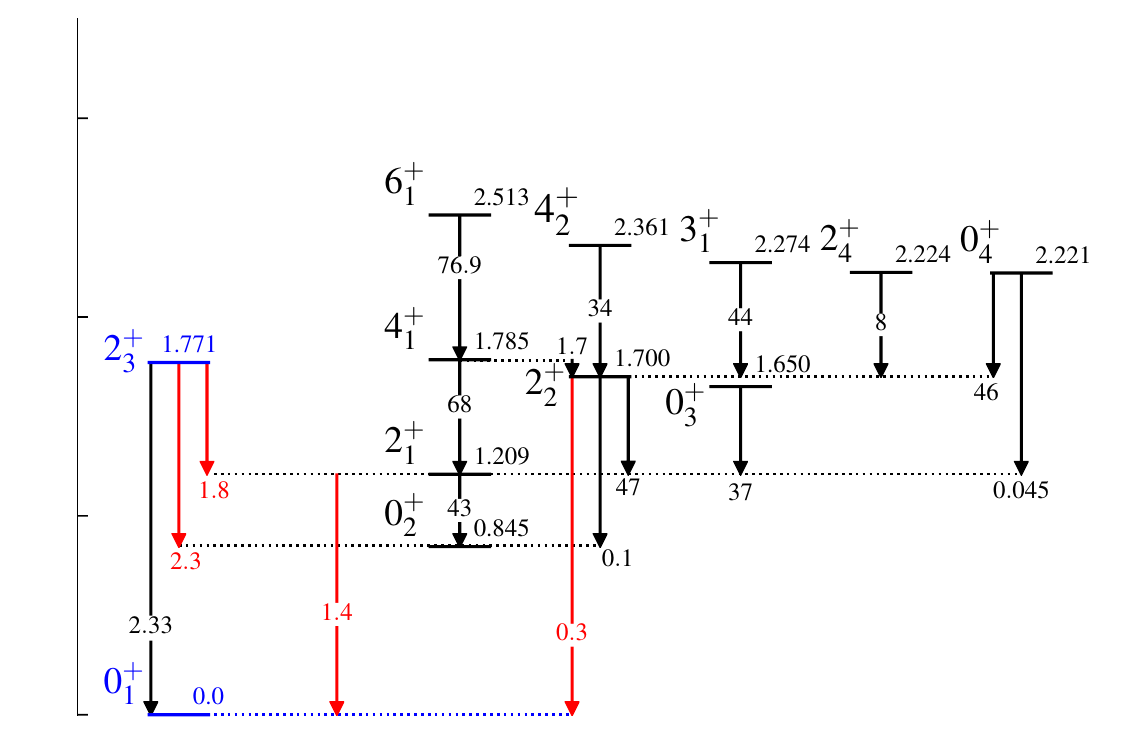}
\put (70,60) {\large (b) {\bf $^{98}$Zr calc}}
\end{overpic}\\
\\
\begin{overpic}[width=0.98\linewidth]{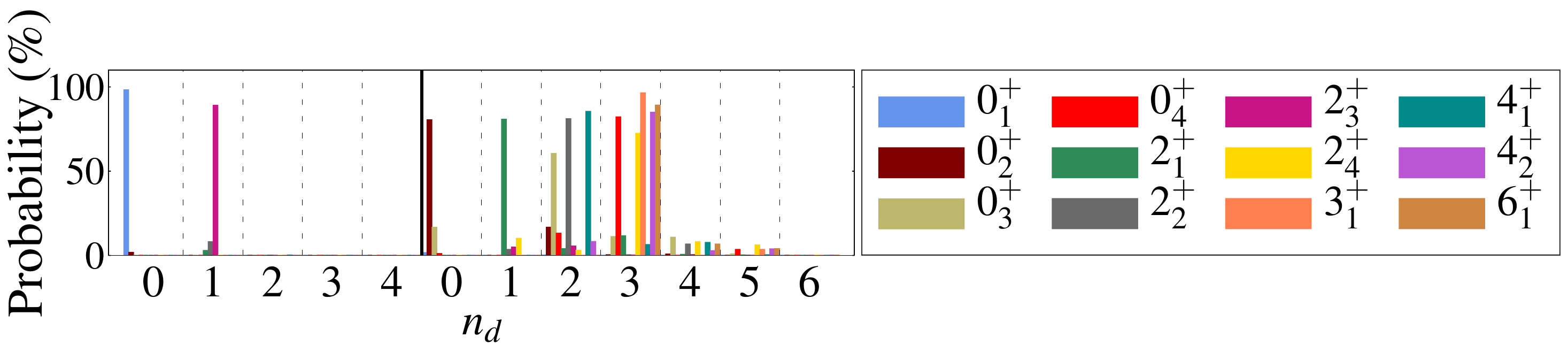}
\put (50,15) {\large (c)}
\put (10,20) {\small Conf.~($A$)}
\put (34,20) {\small Conf.~($B$)}
\end{overpic}
\caption{
\small
(a)~Experimental and (b)~calculated energy levels in MeV
and $E2$ transition rates in
Weisskopf units (W.u.) for $^{98}$Zr.
(c)~U(5) $n_d$-decomposition, Eq.~(\ref{eq:decomp-u5}),
for eigenstates of the
IBM-CM Hamiltonian, Eq.~(\ref{eq:ham_ab}),
assigned to the normal ($A$) and intruder ($B$)
configurations. Adapted from~\cite{Gav2022}.
\label{fig3:98-102Zr-scheme}}
\end{figure*}
\begin{figure}[t]
\centering
\begin{overpic}[width=0.7\linewidth]{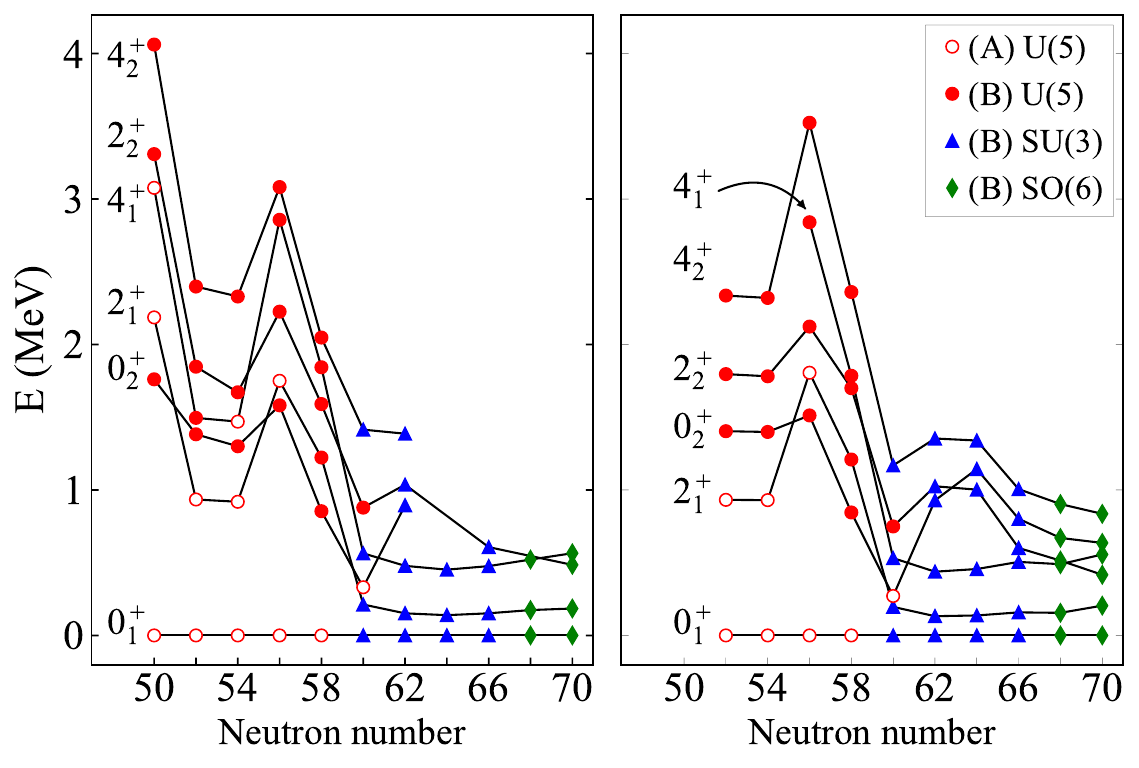}
\put (33.5,61) {\large (a) {\bf Exp}}
\put (56,61) {\large (b) {\bf Calc}}
\end{overpic}
\caption{
\small
Comparison between (a)~experimental and
(b)~calculated energy levels
$0_{1}^{+},2_{1}^{+},4_{1}^{+},0_{2}^{+},2_{2}^{+},4_{2}^{+}$
in the Zr isotopes.
Empty (filled) symbols indicate a state dominated by
the normal $A$ configuration (intruder $B$ configuration),
with assignments based on Eq.~(\ref{Prob-ab}).
The shape of the symbol [$\circ,\triangle,\diamondsuit$],
indicates the closest dynamical symmetry
[U(5), SU(3), SO(6)] to the state considered,
based on Eq.~(\ref{eq:decomp-ds}).
Adapted from~\cite{Gav2019}.
\label{fig4:levels}}
\end{figure}
\subsection{IBM-CM in the Zirconium chain of isotopes}

\hspace{12pt}
The IBM-CM framework
has been used
in~\cite{Gav2019,Gav2020,Gav2022}
to calculate spectral properties of low-lying
states in $^{92-110}$Zr.
In a shell model approach, 
$_{40}^{90}$Zr is taken as a core and valence
neutrons in the 50--82 major shell.
The normal $A$ configuration corresponds to having
no active protons above $Z\!=\!40$ sub-shell gap
and the intruder $B$ configuration corresponds
to two-proton excitation from below to above
this gap, creating 2p-2h states.
In the algebraic approach, the model space consists of
$[N]\oplus[N\!+\!2]$ boson spaces. Here
$N=N_{\pi}+N_{\nu}$ with $N_{\pi}\!=\!0$ ($N_{\pi}\!=\!2)$
proton bosons for the $A$ ($B$) configuration and
$N_{\nu}$ neutron bosons whose number is
determined by the usual boson-counting.
The two configurations relevant for $^{98}$Zr
are shown schematically in
Fig.~\ref{fig2:zr-98-shell}.
For a given isotope and choice of model space,
the parameters of the Hamiltonian and transition operator
are determined by a fit
to the spectra and $E2$ transitions in a manner
described in Ref.~\cite{Gav2022}. Apart from some fluctuations due
to the subshell closure at neutron number 56
(the filling of the $\nu2d_{5/2}$ orbital),
the values of the parameters, shown in Fig.~3 and Table~V
of~\cite{Gav2022}, are 
a smooth function of neutron number and, in some cases, a constant.
A notable exception is the sharp decrease by 1~MeV of
the energy off-set parameter $\Delta$~(\ref{eq:ham_b})
beyond neutron number 56,
reflecting the effects of the proton-neutron residual
interaction~\cite{FedPit79,HeyCas85}.
The parameter $\omega$ of the mixing term~(\ref{eq:mixing}),
is constant except for $^{92,94}$Zr where the normal configuration space
is small ($N\!=\!1,2$).

\hspace{12pt}
As an example of the quantum analysis for
individual nuclei, we present in Fig.~3 the experimental
and calculated spectra and $E2$ rates for low-lying
states in $^{98}$Zr
and their U(5) $n_d$-decomposition.
The assignment of states as normal or
intruder is based on their measured
$E2$ decays when available, or on their calculated
probabilities, $a^2$ and $b^2$, of Eq.~(\ref{Prob-ab}).
The $0^+_1,$ and $2^+_3$ states are assigned to the
normal $A$ configuration. As shown in Fig~3(c), 
the $n_d$-probability of their normal
part, $P^{(N,L)}_{nd}$ of Eq.~(\ref{eq:decomp-u5}), exhibits
a single dominant $n_d$-component ($n_d\approx0,1$),
respectively, which
identifies them as spherical type of states.
The remaining states in Fig.~3(b) are assigned to
the intruder $B$ configuration and exhibit a spectra
of perturbed $n_d$ multiplets. The $n_d$-probability
of their intruder part,
$P^{(N+2,L)}_{nd}$ of Eq.~(\ref{eq:decomp-u5}),
shows a slightly
reduced dominance of single components with
$n_d\approx0,1,2,3$ and a small spread in $n_d$.
These observations suggest that these states are
quasi-spherical or weakly deformed, an interpretation
supported by the measured $E2$ rates.

\hspace{12pt}
A similar analysis can be done for all
$^{92-110}$Zr isotopes. 
In Fig.~\ref{fig4:levels} we show a comparison between
experimental and calculated levels along with
assignments to configurations
and to the closest dynamical symmetry for each state.
In the region between neutron number 50 and 56,
one observes two weakly coupled structures:
a spherical $A$ configuration
(seniority-like)
and a weakly deformed
$B$ configuration, as evidenced by the ratio
$R^{(A)}_{4/2}\cong 1.6 $ and  $R^{(B)}_{4/2} \cong 2.3$.
From neutron number 58, there is a pronounced drop in
energy for the states of the $B$ configuration
and at 60, the two configurations exchange their role,
indicating a Type~II QPT. 
At this stage, the intruder $B$ configuration appears
to be at the critical point of a U(5)-SU(3) Type~I QPT,
as evidenced by the low value of the
energy of the first excited $0^+$
state of this configuration.
Beyond neutron number 60, the intruder
$B$ configuration is strongly deformed, as evidenced
by the small value of $E(2^{+}_2)$
and by the ratio $R^{(B)}_{4/2}\!=\!3.24$ in $^{104}$Zr.
At still larger neutron number 66,
the ground state band becomes $\gamma $-unstable,
as evidenced by the close energy of the states
$2_{2}^{+}$ and $4_{1}^{+}$ in $^{106}$Zr and $^{110}$Zr,
a signature of the SO(6) symmetry. 
In this region, the ground ($B$) configuration
shows spectral features of a~crossover from SU(3) to SO(6). 

\newpage
\subsection{Intertwined QPTs in the Zr chain}

\hspace{12pt}
The above spectral analysis suggests a situation of
coexisting
Type~I and Type~II QPTs, which is the defining property
of intertwined QPTs. In order to understand
the nature of these phase transitions, it is instructive
to examine the behavior of the order parameters,
the configuration content and symmetry properties of
the evolving states.

\hspace{12pt}
Fig.~5(a) shows the evolution with neutron number
of the order parameters,
$\braket{\hat{n}_d}_{0^{+}_1}$, $\braket{\hat{n}_d}_A$,
$\braket{\hat{n}_d}_B$,
Eq.~(\ref{eq:order-param-cm}),
of the ground state wave function,
$\ket{\Psi; L\!=\!0^{+}_1}$ and its
$\Psi_A$, $\Psi_B$ components,
normalized by the respective boson numbers:
$\braket{\hat N}_{0^+_1}\!=\! a^2N + b^2(N+2)$,
$\braket{\hat N}_A \!=\! N, \braket{\hat N}_B \!=\! N+2$.
The $A$ configuration is seen to be spherical for all
neutron numbers considered. 
In contrast, the $B$ configuration is weakly-deformed
for neutron number 52-58. One can see here clearly
a jump between 58 and 60 from the $A$-
to the $B$ configuration, indicating a 
Type~II QPT, a further increase at 60-64,
indicating a U(5)-SU(3) Type~I QPT and,
finally, there is a decrease at 66,
indicating a crossover from SU(3) to SO(6).
$\braket{\hat{n}_d}_{0^{+}_1}$ is close to
$\braket{\hat{n}_d}_A$ for 52-58 and
coincides with $\braket{\hat{n}_d}_B$ at 60 and above,
consistent with a high degree of purity with respect 
to configuration-mixing.

\hspace{12pt}
Fig.~5(b) shows the probability $b^2$ of the $\Psi_B$
component, Eq.~(\ref{Prob-ab}), in the wave functions
of the ground state ($0^+_1$) and first-excited
state ($2^+_1$). The rapid change in structure
of $0^+_1$ from the normal
$A$ configuration in $^{92-98}$Zr
(small $b^2$) to the intruder
$B$ configuration in $^{100-110}$Zr
(large $b^2$) is clearly evident,
signaling a Type II QPT.
The configuration change appears sooner in $2^+_1$,
which changes to the
$B$ configuration already in $^{98}$Zr.
Outside a narrow region near neutron number 60, where
the crossing occurs, the two configurations are weakly
mixed and the states retain a high level of purity.

\hspace{12pt}
The shape-evolution within the $B$ configuration is
reflected in the change of symmetries in the
lowest $0^{+}_B$ state of this configuration. As shown
in Fig.~5(c), the probability
of the U(5) irrep $n_d=0$, Eq.~(\ref{eq:decomp-u5}),
is large ($\sim90\%$) for neutron number 52--58,
indicating a spherical type of state.
At 60 and beyond, it drops drastically,
indicating appreciable $n_d$-mixing and a transition
to a deformed state. This behavior is consistent
with the observed rise in probability
of the SU(3) irrep $(\lambda,\mu)=(2N+4,0)$,
Eq.~(\ref{eq:decomp-su3}),
for neutron number 60-64,
signaling a U(5)-SU(3) Type I QPT.
From 66 and beyond one sees a dominant
probability of the SO(6) irrep $\sigma=N+2$,
Eq.~(\ref{eq:decomp-so6}), suggesting a crossover
from SU(3) to SO(6). A very similar trend is observed
for the $2^+_B$ state.
The findings of Fig.~5
support the claim for the occurrence
of two configurations in Zr isotopes
that are weakly mixed and interchange
their roles in the ground state while their individual
shapes evolve gradually with neutron number,
\textit{i.e.} intertwined Type I and II QPTs.

\subsection{Evolution of observables along the Zr chain}

\hspace{12pt}
The above conclusions are stressed by an analysis of
other observables.
In particular, the $B(E2)$ values
for $2^{+}\to 0^{+}$ transitions, shown in Fig.~6(a),
follow the same trends as the 
respective order parameters, seen in Fig.~5(a).
The isotope shift
$\Delta\braket{\hat r^2}_{0^+_1}$, shown in Fig.~6(b),
increases at the transition point and decreases
afterwards, in accord with the expected
behaviour of a first-order Type~I QPT.
(In the large $N$ limit, this quantity, proportional
to the derivative of the order parameter
$\braket{\hat{n}_d}_{0^{+}_1}$,
diverges at the critical-point).
The two-neutron separation energy $S_{2n}$,
shown in Fig.~6(c),
is a straight line for neutron number 52-56, 
as the ground state is spherical (seniority-like)
$A$ configuration. After 56, it first goes down due
to the subshell closure at~56, then it flattens
as expected from a first-order Type~I QPT.
After 62, it goes down again due to the increase of
deformation and finally it flattens as expected
from a crossover from SU(3) to SO(6).
Lastly, the magnetic moments of the $2^+_1$ states
in Zr isotopes,
shown in Fig.~6(d), exhibit a jump from small (or negative)
single-particle values for neutron numbers 52--56,
to large collective (rigid rotor) values for 58-70.
\definecolor{GreenN}{rgb}{0,0.5,0}
\definecolor{GrayN}{rgb}{0.8,0.8,0.9003921568627451}
\begin{figure}[]
\centering
\begin{overpic}[width=0.5\linewidth]{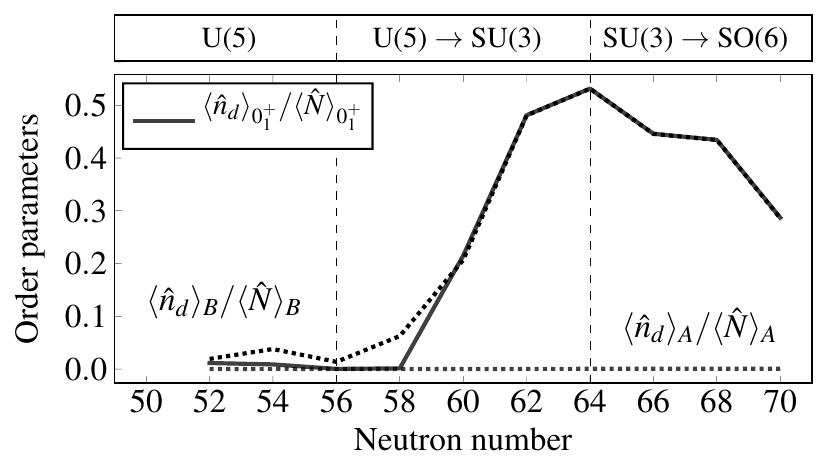}
\put (18,32) {\large(a)}
\end{overpic}
\begin{overpic}[width=0.49\linewidth]{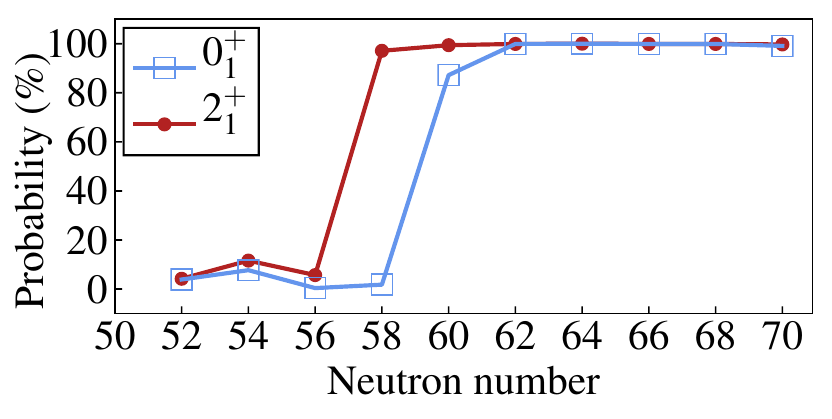}
\put (88,32) {\large(b)}
\end{overpic}
\begin{overpic}[width=0.8\linewidth]{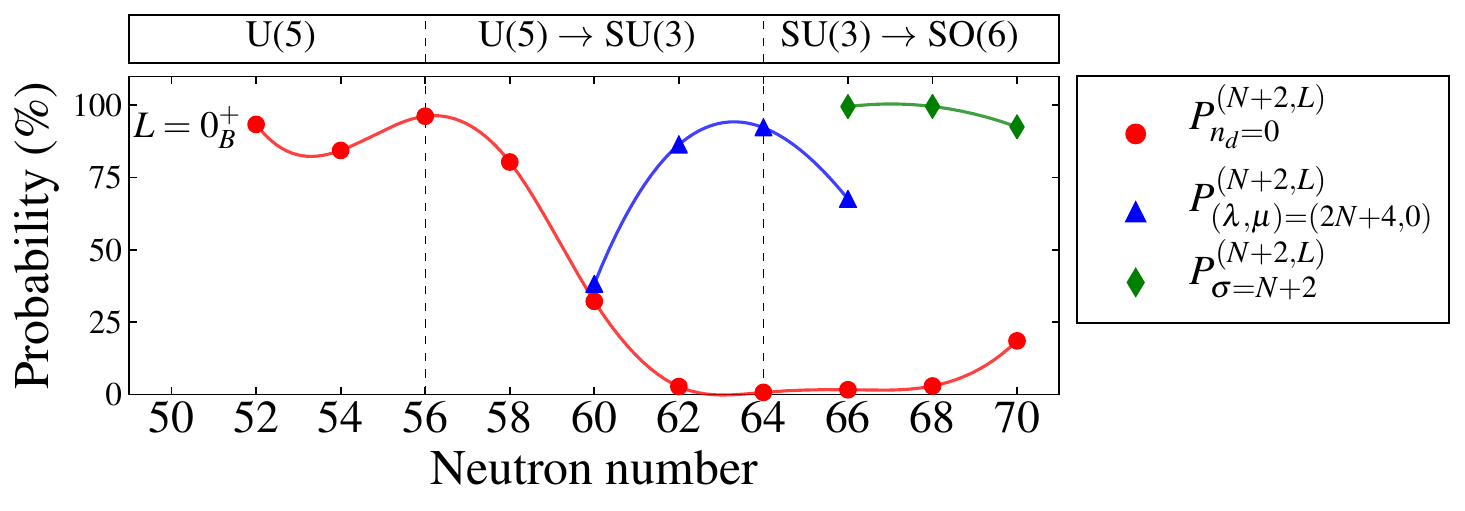}
\put (12,18) {\large(c)}
\end{overpic}
\caption{
\small
Evolution of structure along the Zr chain.
(a)~Order parameters: $\braket{\hat n_d}_{0^+_1}$
(solid line) and
$\braket{\hat n_d}_{A},\,\braket{\hat n_d}_{B}$
(dotted lines), defined in Section~2.4, 
normalized by the respective
boson numbers.
(b)~Percentage of the wave functions within the
intruder $B$ configuration [the $b^2$ probability of
Eq.~\eqref{Prob-ab}], for the ground ($0^+_1$)
and excited ($2^+_1$) states.
(c)~Symmetry properties of the lowest $0^+_B$ state
in the intruder $B$ configuration.
Shown are the probabilities of selected
U(5)~(${\color{red}\CIRCLE}$),
SU(3)~(${\color{blue}\blacklozenge}$),
and SO(6)~(${\color{GreenN}\blacktriangle}$) irreps
in $0^+_B$,
obtained from Eq.~(\ref{eq:decomp-ds}).
\label{fig5:mixing}}
\end{figure}
\begin{figure}[t!]
\centering
\begin{overpic}[width=0.55\linewidth]{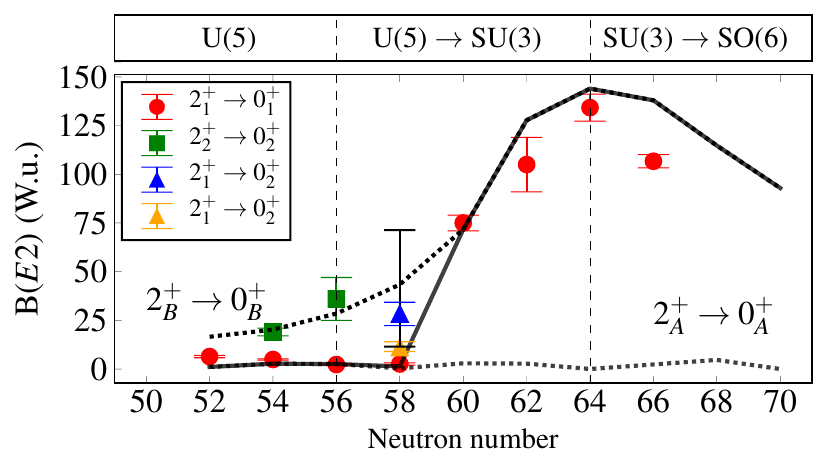}
\put (50,40) {\large(a)}
\end{overpic}
\begin{overpic}[width=0.44\linewidth]{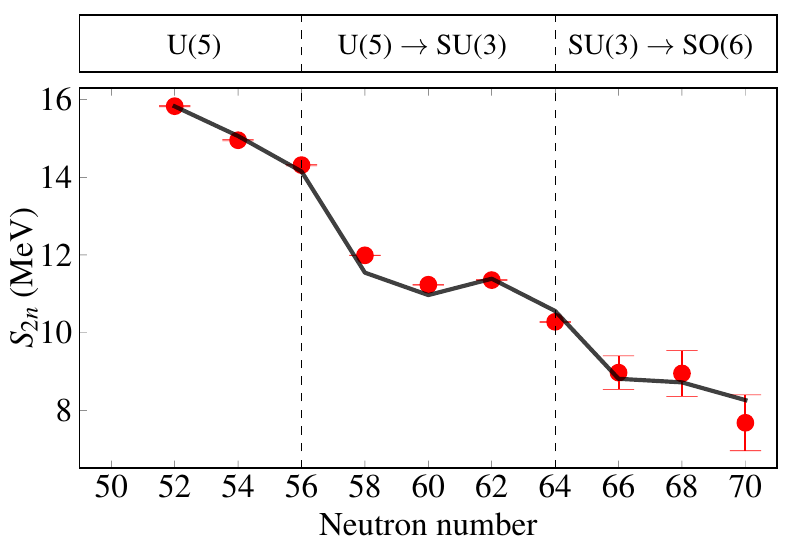}
\put (82,49) {\large(c)}
\end{overpic}\\
\begin{overpic}[width=0.542\linewidth]{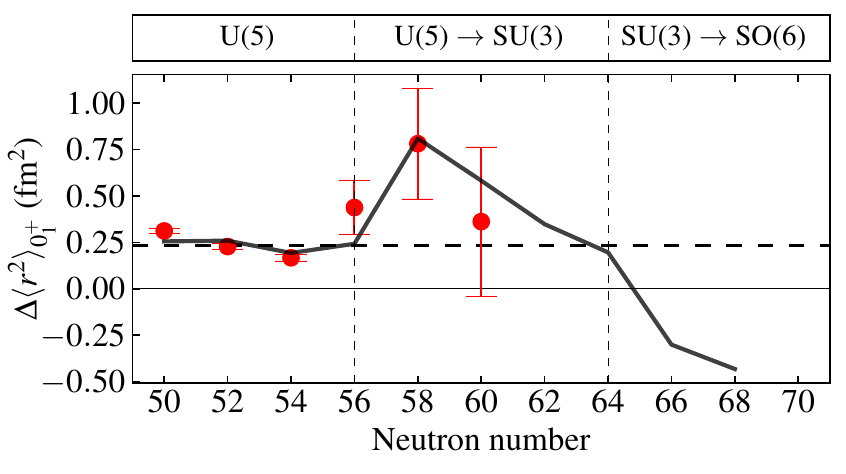}
\put (82,33) {\large(b)}
\end{overpic}
\begin{overpic}[width=0.45\linewidth]{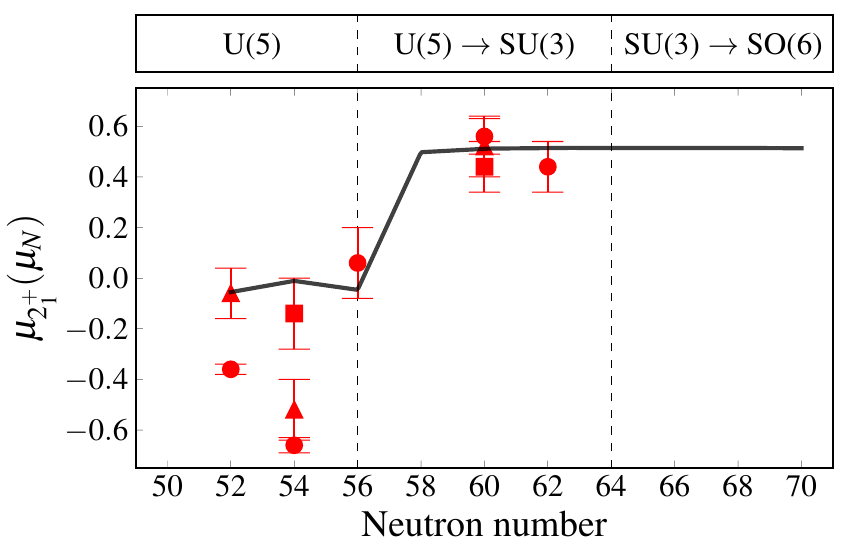}
\put (82,38) {\large(d)}
\end{overpic}
\caption{
\small
Evolution of observables along the Zr chain.
Symbols (solid lines) denote experimental data (calculated
results).
(a)~$B(E2;2^{+}\to 0^{+})$ in W.u.
(b)~Isotope shift $\Delta\braket{\hat{r}^{2}}_{0^{+}_1}$
in fm$^{2}$.
(c)~Two-neutron separation energies $S_{2n}$ in MeV.
(d)~Magnetic moments $\mu_{2^+_1}$ in units of
nuclear magneton ($\mu_N$).
\label{fig6:be2}}
\end{figure}
\begin{figure*}[t!]
\centering
\begin{overpic}[width=0.19\linewidth]{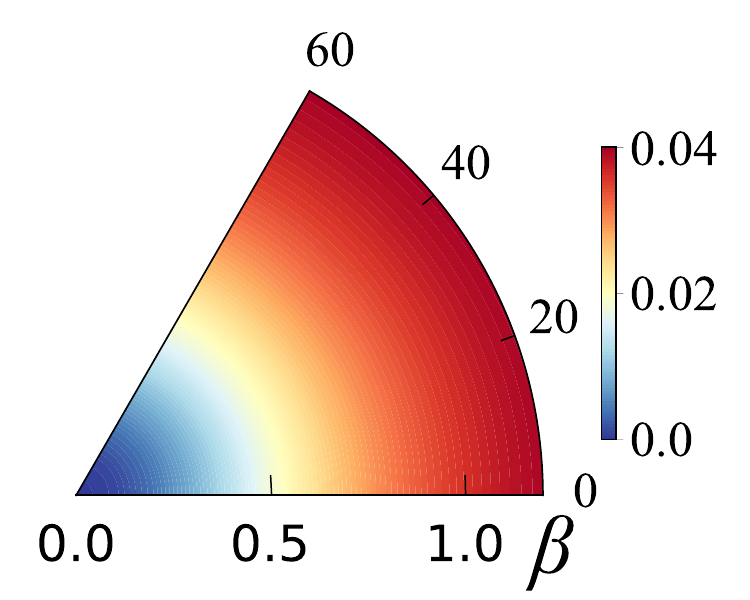}
\put (0,50) {\large $^{92}$Zr}
\end{overpic}
\begin{overpic}[width=0.19\linewidth]{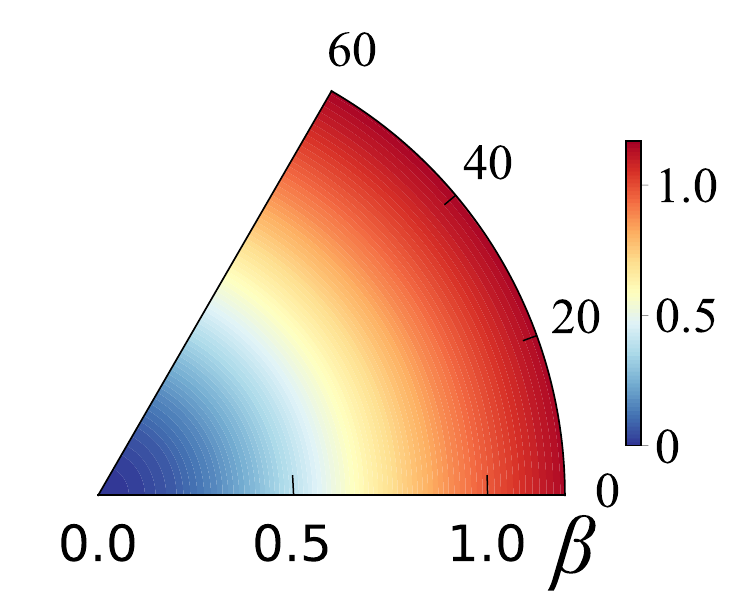}
\put (0,50) {\large $^{94}$Zr}
\end{overpic}
\begin{overpic}[width=0.19\linewidth]{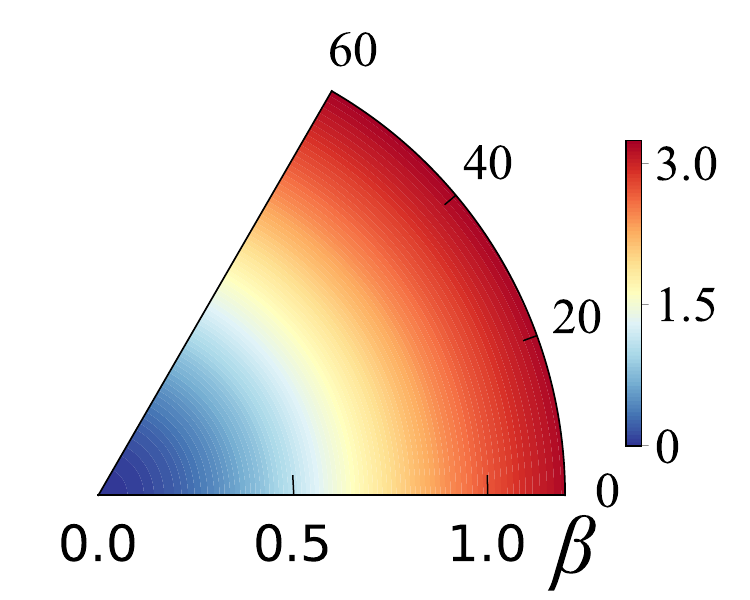}
\put (0,50) {\large $^{96}$Zr}
\end{overpic}
\begin{overpic}[width=0.19\linewidth]{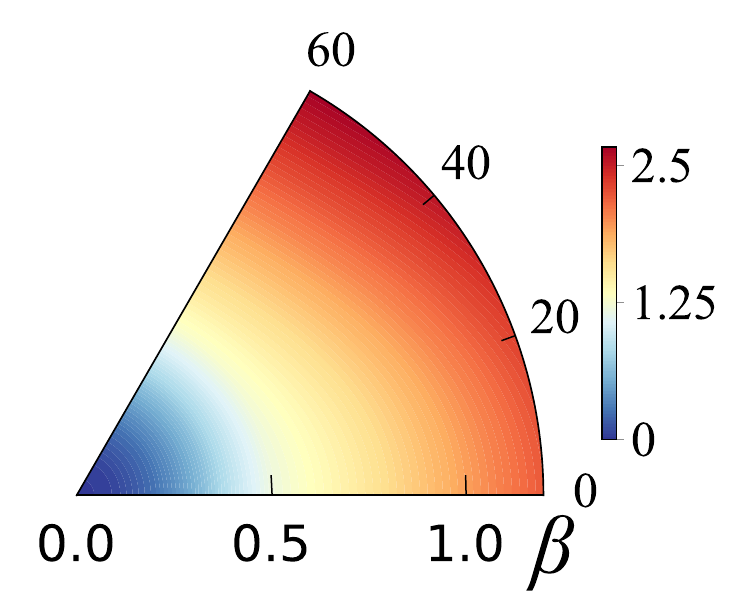}
\put (0,50) {\large $^{98}$Zr}
\end{overpic}
\begin{overpic}[width=0.19\linewidth]{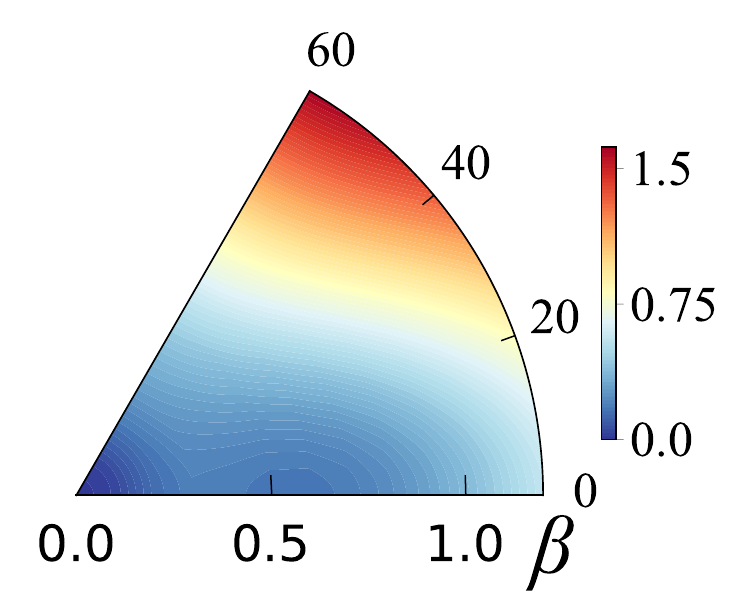}
\put (0,50) {\large $^{100}$Zr}
\end{overpic} \\ 
\begin{overpic}[width=0.19\linewidth]{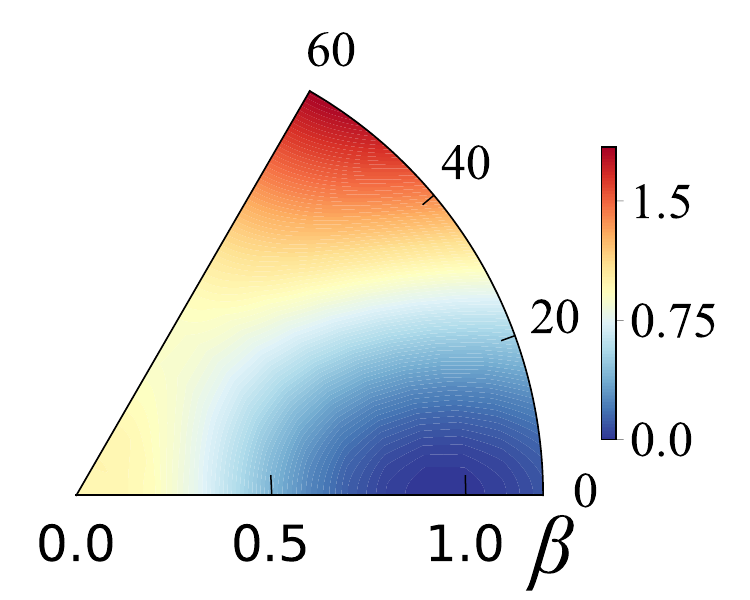}
\put (0,50) {\large $^{102}$Zr}
\end{overpic}
\begin{overpic}[width=0.19\linewidth]{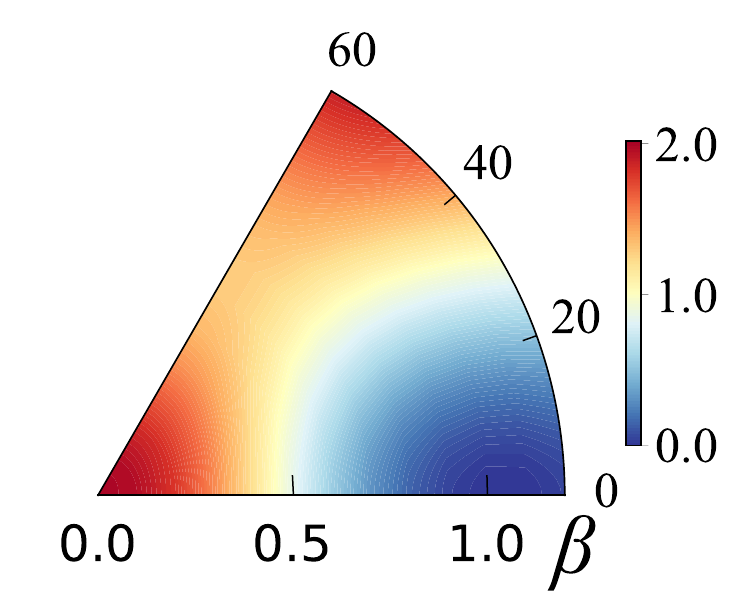}
\put (0,50) {\large $^{104}$Zr}
\end{overpic}
\begin{overpic}[width=0.19\linewidth]{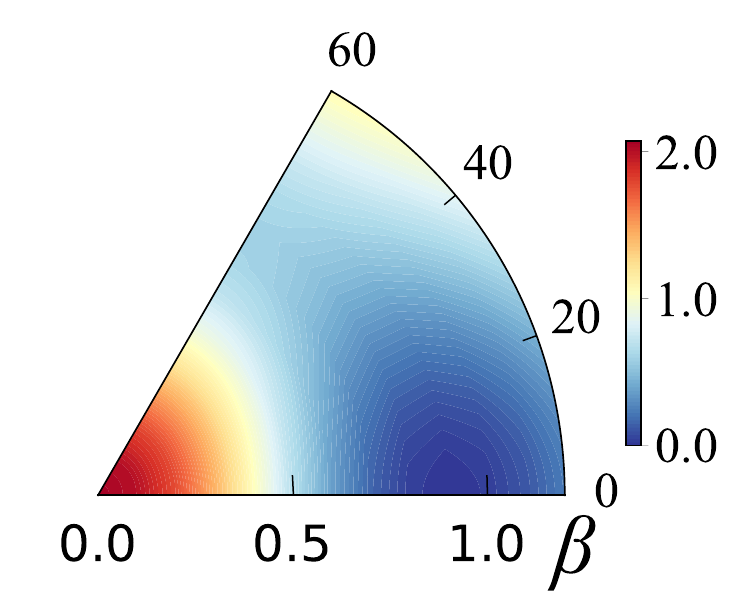}
\put (0,50) {\large $^{106}$Zr}
\end{overpic}
\begin{overpic}[width=0.19\linewidth]{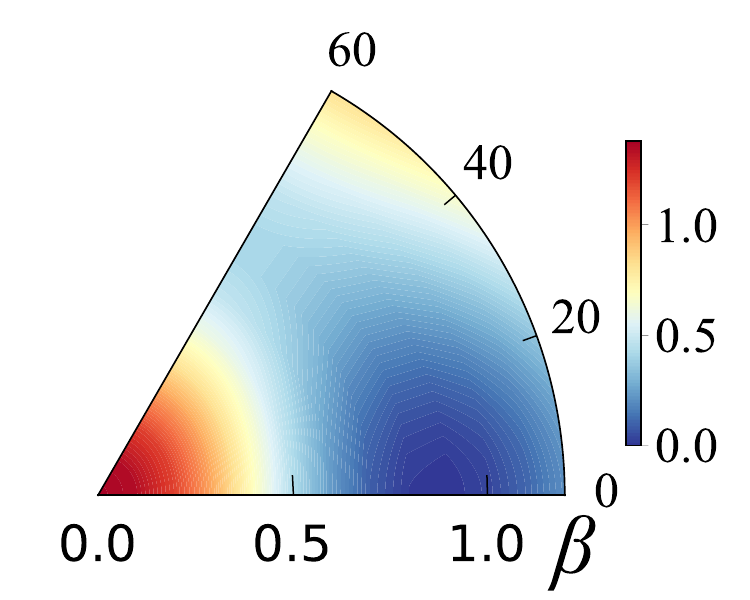}
\put (0,50) {\large $^{108}$Zr}
\end{overpic}
\begin{overpic}[width=0.19\linewidth]{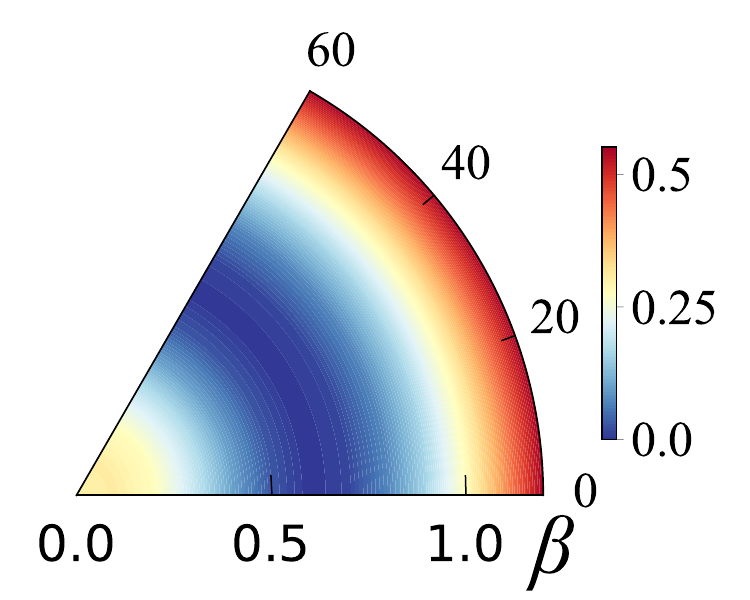}
\put (0,50) {\large $^{110}$Zr}
\end{overpic}
\caption{\label{fig7:Eminus}
\small
Contour plots in the $(\beta ,\gamma )$ plane of the
lowest eigen-potential surface, $E_{-}(\beta ,\gamma )$
mentioned in Section~2.4,
for the $^{92-110}$Zr isotopes.
$\beta$ is a radial coordinate and the angle $\gamma$ is
in degrees. Adapted from~\cite{Gav2019}.}
\end{figure*}

\subsection{Classical analysis}

\hspace{12pt}
One of the main advantages of the algebraic approach is
that one can do both a quantum and a classical analysis. 
In Fig.~7, we show the calculated 
lowest eigen-potential $E_{-}(\beta ,\gamma )$,
obtained by diagonalizing the matrix of
Eq.~(\ref{eq:surface-mat}).
These classical potentials confirm the quantum results,
as they show a transition from spherical ($^{92-98}$Zr),
to a flat-bottomed potential at $^{100}$Zr,
to prolate axially-deformed ($^{102-104}$Zr), and finally to 
$\gamma$-unstable ($^{106-110}$Zr).

\newpage
\section{Algebraic approach to quantum phase
  transitions in Bose-Fermi systems}

\hspace{12pt}
The study of QPTs in a Bose-Fermi system is more complex
due to the presence of both collective (bosonic)
and single-particle (fermionic) degrees of freedom
with different statistics. As a representative for
the algebraic approach to such systems, we consider 
the interacting boson-fermion model (IBFM)~\cite{IBFMBook},
a model of odd-even nuclei phrased in terms
of correlated pairs with $L=0,2$ ($s,d$ bosons) 
and unpaired particles with angular momentum~$j$
($j$~fermions). The model is based on a
$U_{b}(6)\otimes U_{f}(\Omega)$ spectrum generating
algebra, where $U_{b}(6)$ is the
bosonic algebra of the IBM discussed in Section~2.1
and $U_{f}(\Omega)$ is the fermion algebra with 
$\Omega$ the dimension of the single-fermion space.
Here we focus the discussion to
the case of a single-$j$ fermion for which $\Omega=2j+1$.

\hspace{12pt}
The IBFM Hamiltonian contains
boson- fermion- and boson-fermion parts,
\ba
\label{H-IBFM}
\hat H = \hat H_{\rm b} + \hat H_{\rm f}
+ \hat V_{\rm bf} ~,
\ea
where,
\bsub
\label{H-b-f-bf}
\ba
\hat H_{\rm b}(\epsilon_d,\kappa,\chi) &=&
  \epsilon_d\,
  \hat n_d + \kappa\,
  \hat Q_\chi \cdot \hat Q_\chi ~,\\
  \hat H_{\rm f}(\epsilon_j) &=&
  \epsilon_j\,\hat{n}_j ~,\\
\hat V_{\rm bf}(A,\Gamma,\Lambda) &=&
V_{\rm bf}^{\rm MON} + \hat{V}_{\rm bf}^{\rm QUAD}
+ \hat{V}_{\rm bf}^{\rm EXC} ~.
\ea
\esub
$\hat{H}_{\rm b}$
is the IBM 
Hamiltonian, taken to be as in Eq.~(\ref{eq:ham-q}),
describing the dynamics of the even-even core.
$\hat{H}_{\rm f}$ involves the fermion
number operator $\hat{n}_j=\sum_{m}a^{\dag}_{j,m}a_{j,m}$ and
$\hat{V}_{\rm bf}$ is composed of monopole, quadrupole and
exchange terms given by,
\bsub
\label{V-bf}
\ba
V_{\rm bf}^{\rm MON} &=&
A\,\hat{n}_d\,\hat{n}_j ~,\\
\hat{V}_{\rm bf}^{\rm QUAD} &=& 
\Gamma\,\hat{Q}_{\chi}\cdot
( a_{j}^{\dag }\times \tilde{a}_{j} )^{(2)} ~,\\
\hat{V}_{\rm bf}^{\rm EXC} &=& 
\Lambda
\sqrt{2j+1}:[ ( d^{\dag }\, \tilde{a}_{j})^{(j)}\times
  ( \tilde{d}\, a_{j}^{\dag })^{(j)}]^{(0)}: ~, 
\ea
\esub
where $\tilde{a}_{j,m} \!=\! (-1)^{j-m}a_{j,-m}$
and  $:\,:$ denotes normal ordering.

\subsection{Type I QPT}

\hspace{12pt}
In Type~I QPTs, the coupling constants of the
IBFM Hamiltonian,
Eqs.~(\ref{H-IBFM})-(\ref{V-bf}),
serve as control parameters. As discussed in Section~2.3,
the boson part $\hat H_{\rm b}(\epsilon_d,\kappa,\chi)$
interpolates between the DS limits of the IBM and
describes a shape-phase transition in the
even-even core.
The boson-fermion terms in
$\hat{V}_{\rm bf}(A,\Gamma,\Delta)$ control
the coupling of the odd fermion to the boson core.
IBFM Hamiltonians of the above form have been widely
used to study Type~I QPTs in odd-even
nuclei~\cite{IBFMBook, ScholtenBlasi1982,Alonso2005,
Boyukata2021,Petrellis2011a,Nomura2020}.

\hspace{12pt}
For studying the effect of the fermion on QPTs in bosonic
systems and interpreting the structural changes taking
place, it is instructive to consider known coupling
schemes of the odd fermion to different core shapes. 
If the shape of the core is spherical,
the weak coupling scheme is appropriate, where
the wave functions are written as,
\ba
\ket{\Psi;[N],L,j;J} =
\ket{N,n_d,\tau,n_{\Delta},(L\otimes j)J} \qquad
  J=|L-j|,\, |L-j|+1,\ldots, (L+j) ~.
\label{weak-coupl}
\ea  
Here U(5) basis states
$\ket{N,n_d,\tau,n_{\Delta},L}$, Eq.~(\ref{U5-ds}),
are coupled to the fermion state $\ket{j}$, to form
a multiplet of states with total angular
momentum $J$ in the range of values specified above.

\hspace{12pt}
If the core has an axially-deformed prolate shape,
the strong coupling scheme is appropriate, 
with wave functions of the form,
\ba
\ket{\Psi;[N],j,K;J} =
\sqrt{\frac{2}{2J+1}}\sum_{L=0,2,4,\ldots}
\sqrt{2L+1}(L,0;j,K|J,K)\,
\ket{N,(2N,0),K_b=0,(L\otimes j)J} ~,
\label{strong-coupl}
\ea
where $J=K,K+1,K+2,\ldots$
and $(\;\; |\;)$ stands for a Clebsch Gordan coefficient.
Here states of the SU(3) ground band
$\ket{N,(2N,0),K_b=0,L}$, Eq.~(\ref{SU3-ds})
with $L\geq 0$ even,
are coupled to the angular momentum $j$ of the odd
fermion to form rotational $K$-bands
with angular momentum projection $K=j,\,j-1,\ldots,1/2$,
along the symmetry axis.
The relative order of different $K$-bands is governed by
the strengths $\Gamma$ and $\Lambda$ of the quadrupole and
exchange terms~\cite{Lev88}.
For a general prolate-deformed shape,
away from the SU(3) limit,
the relevant strong-coupling wave functions are
obtained by replacing in Eq.~(\ref{strong-coupl})
the SU(3) basis states by the $L$-states
projected from the intrinsic state
$\ket{\beq,\gaeq=0;N}$, Eq.~(\ref{int-state}),
with the equilibrium deformations.

\subsection{Type~II QPT}

\hspace{12pt}
To allow for core excitations and accommodate several
configurations, the IBFM can be
extended to obtain the interacting boson-fermion model
with configuration mixing
(IBFM-CM)~\cite{gavleviac22,Gav23}.
The Hamiltonian
is composed of boson-, fermion-
and boson-fermion parts,
which now become matrices. For two configurations
($A$, $B$) it has the form
$\hat H = \hat H_{\rm b} + \hat H_{\rm f}
+ \hat V_{\rm bf}$ with,
\begin{equation}
\label{eq:H_bfmcm}
\hat H_{\rm b} = \begin{bmatrix}
  \hat H_{\rm b}^{\rm A}(\xi_{\rm A}) &
\hat{W}_{\rm b}(\omega)\\
\hat{W}_{\rm b}(\omega) 
& \hat H_{\rm b}^{\rm B}(\xi_{\rm B})
\end{bmatrix} \;\;,\;\;
\hat H_{\rm f} = \begin{bmatrix}
\epsilon^{\rm A}_j\, \hat n_j & 0 \\
0 & \epsilon^{\rm B}_j\, \hat n_j
\end{bmatrix} \;\;,\;\;
\hat V_{\rm bf} = \begin{bmatrix}
  \hat V^{\rm A}_{\rm bf}(\zeta_{\rm A}) &
 \hat{W}_{\rm bf}(\omega_j)\\
\hat{W}_{\rm bf}(\omega_j) &
\hat V^{\rm B}_{\rm bf}(\zeta_{\rm B})
\end{bmatrix} ~.
\end{equation}
$\hat H_{\rm b}$ coincides with the IBM-CM Hamiltonian,
discussed in Section~2.4,
with entries given in Eq.~(\ref{eq:ham_ab}). It~acts
in the boson spaces $[N]\oplus[N\!+\!2]$, representing
the normal ($A$) and intruder ($B$) configurations.
$\hat H_{\rm f}$ involves the one-body fermion number
operator for each configuration.
$\hat V_{\rm bf}$ involves monopole, quadrupole
and exchange two-body terms, Eq.~(\ref{V-bf}),
for each configuration and a mixing term,
\ba
\hat{W}_{\rm bf} = \omega_j\, \hat n_j
[(d^{\dag}d^{\dag})^{(0)} + (s^{\dag})^2 + \text{H.c.}] ~.
\label{Wbf}
\ea
For simplicity, the parameters of
$\hat{H}_{\rm f}$ and $\hat{V}_{\rm  bf}$
are taken to be the same in both configurations,
{\it i.e.}, $\epsilon^{\rm A}_j = \epsilon^{\rm B}_j
= \epsilon_j$ and $\zeta_A=\zeta_B=(A,\Gamma,\Lambda)$,
and $\omega_j=0$ since $\hat{W}_{\rm bf}$ coincides with
$\hat{W}_{\rm b}$ for a single-$j$ fermion.

\hspace{12pt}
The eigenstates $\ket{\Psi;j,J}$
of the IBFM-CM Hamiltonian, Eq.~(\ref{eq:H_bfmcm}),
are linear combinations of wave functions $\Psi_{\rm A}$ 
and $\Psi_{\rm B}$, involving bosonic basis states in the 
two spaces $\ket{[N],\alpha,L}$ and
$\ket{[N\!+\!2],\alpha,L}$, coupled to a $j$-fermion.
Here $\alpha$ denotes additional quantum~numbers of the
dynamical symmetry chain, Eq.~(\ref{IBMchains}).
The boson ($L$) and fermion ($j$) angular momenta are 
coupled to $J$,
\ba
\ket{\Psi;j,J} &=&
a\ket{\Psi_A,[N],j;J} + b\ket{\Psi_B,[N\!+\!2],j;J}
\nonumber\\[2mm]
&=&
\sum_{\alpha,L}C^{(N,J)}_{\alpha,L,j}
  \ket{N,\alpha,(L\otimes j)J}
+ \sum_{\alpha,L}C^{(N+2,J)}_{\alpha,L,j}
\ket{N+2,\alpha,(L\otimes j)J} ~.
\ea
The probability of normal-intruder mixing is given by,
\begin{equation}
\label{Prob-ab-cm}
  a^2 =\sum_{\alpha,L}|C^{(N,J)}_{\alpha,L,j}|^2 \;\;\; ,
  \;\;\;
  b^2 = \sum_{\alpha,L}|C^{(N+2,J)}_{\alpha,L,j}|^2
= 1-a^2 ~.
\end{equation}
Operators inducing electromagnetic transitions
contain boson and fermion parts,
{\it e.g.}, the $E2$ operator has the form
$\hat{T}(E2) = e_{\rm A}\,\hat Q^{(N)}_{\chi}
+ e_{\rm B}\,\hat Q^{(N+2)}_{\chi}
+ e_f\,(a^{\dag}_j\times\tilde{a}_j)^{(2)}$,
with effective charges $(e_{\rm A},e_{\rm B},e_f)$.
\begin{figure}[t!]
\centering
\includegraphics[width=0.7\linewidth]{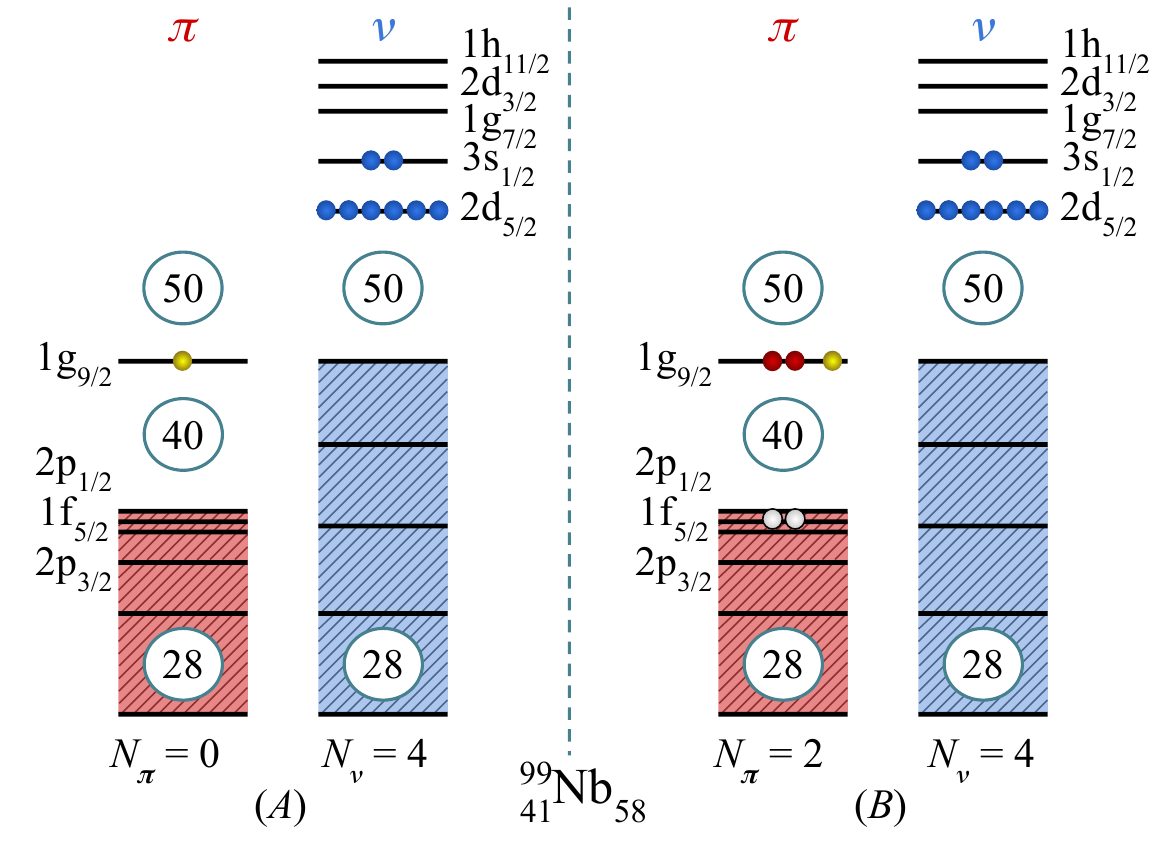}
\caption{
\small
Schematic representation of the two
shell-model configurations ($A$ and $B$) for
$^{99}_{41}$Nb$_{58}$. The corresponding numbers of
proton bosons ($N_{\pi}$) and neutron bosons ($N_{\nu}$),
are listed for each configuration and $N=N_{\pi}+N_{\nu}$.
\label{fig8:nb-99-shell}}
\end{figure}

\subsection{IBFM-CM in the Niobium chain of isotopes}

\hspace{12pt}
The IBFM-CM framework has been applied
in~\cite{gavleviac22,Gav23}
to calculate spectral properties of low-lying states in
the $^{A}_{41}$Nb isotopes with mass number 
\mbox{$A\!=\!\text{93--105}$}.
In a shell model approach,
these isotopes are described by coupling a 
proton to their respective $_{40}$Zr cores with
neutron number 52--64.
The latter involve the normal ($A$)
and intruder ($B$) configurations, as outlined
in Section~2.5.
In the present contribution,
we focus on the positive-parity 
states in the Nb isotopes
(negative parity states are addressed in~\cite{Gav23}).
Such a case reduces to a 
single-$j$ calculation where the IBFM-CM model space
consists of a proton in a $\pi(1g_{9/2})$ orbit
plus $[N]\oplus[N\!+\!2]$ boson spaces.
The two configurations relevant for $^{99}$Nb
are shown schematically in Fig.~8.
The parameters of the IBFM-CM Hamiltonian and transition
operators are determined from a fit in the manner
described in~\cite{gavleviac22,Gav23}.
The parameters of the boson part $\hat H_{\rm b}$
are taken to be the same as in the Zr
calculation~\cite{Gav2022}.
The Bose-Fermi couplings in
$\hat{V}_{\rm bf}(A,\Gamma\Lambda)$
are expressed by occupation probabilities and strengths,
whose values are listed in Table~I
of~\cite{gavleviac22}. The latter are
either constant for the entire chain or segments of it,
and vary smoothly.
\begin{figure}[t!]
\centering
\includegraphics[width=0.7\linewidth]{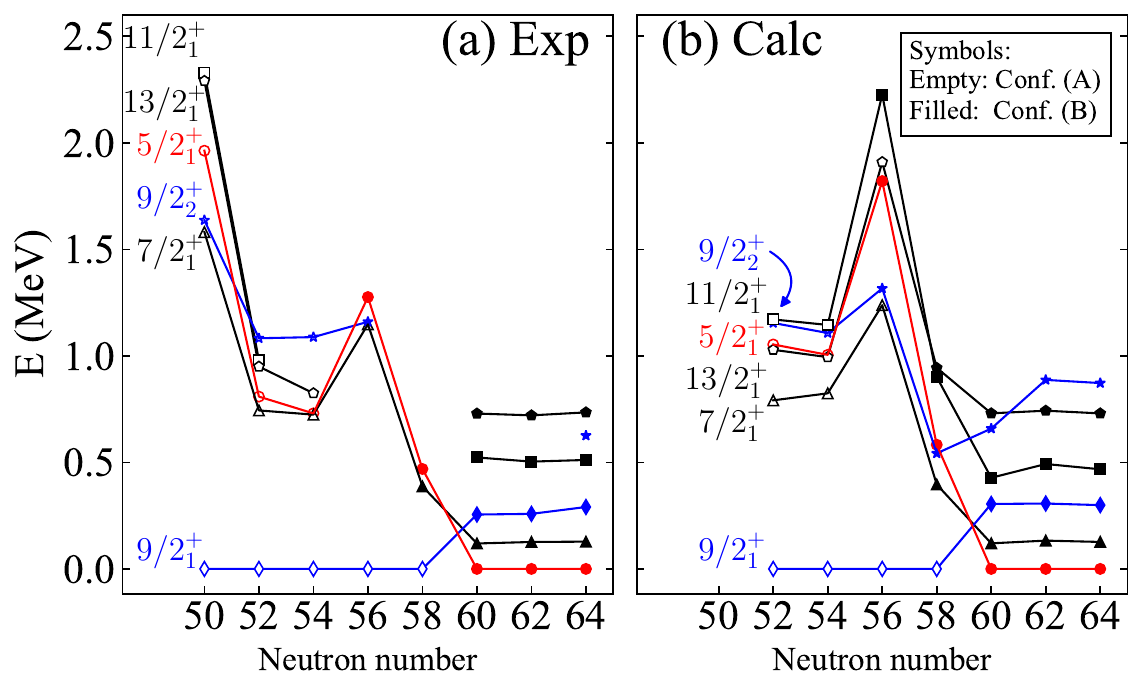}
\caption{
\small
Comparison between
(a)~experimental
and (b)~calculated lowest-energy   
positive-parity levels in Nb isotopes. Empty (filled) 
symbols indicate a state dominated by the normal 
$A$ configuration (intruder $B$ configuration), with   
assignments based on Eq.~(\ref{Prob-ab-cm}).
In particular,   
the $9/2^+_1$ state is in the $A$ ($B$) configuration for  
neutron number 52--58 (60--64) and the $5/2^+_1$ state is  
in the $A$ ($B$) configuration for 52--54 (56--64).
Adapted from~\cite{gavleviac22}.
\label{fig9:energies-p}}
\end{figure}
\begin{figure}[t!]
\centering
\includegraphics[width=0.49\linewidth]{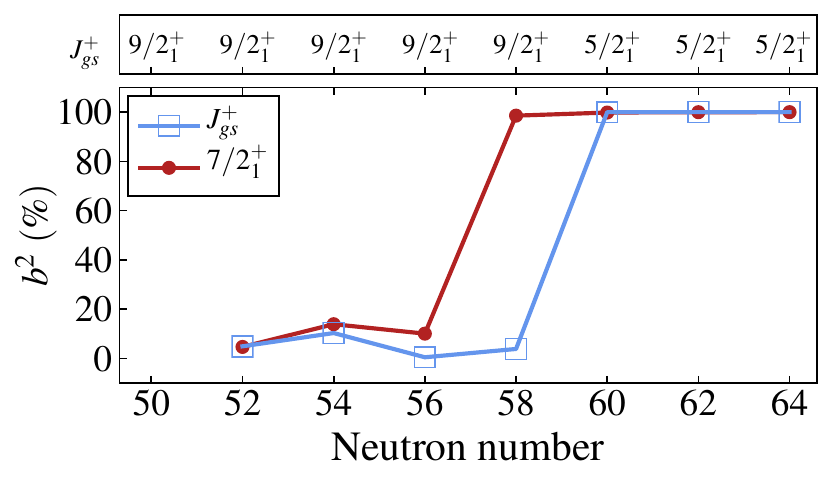}
\caption{
\label{fig10:b2}  
Percentage of the intruder ($B$) 
component [the $b^2$ probability
of Eq.~(\ref{Prob-ab-cm})],
in the ground state ($J^+_{gs}$) and
first-excited positive-parity state ($7/2^+_1$) of 
$^{93-103}$Nb. The values of $J^+_{gs}$ are indicated at 
the top.}
\end{figure}

\hspace{12pt}
Figures 9(a) and 9(b) show the experimental
and calculated levels of selected states
in the Nb isotopes
along with assignments to configurations based on 
Eq.~(\ref{Prob-ab-cm}).
In the region between neutron 
number 50 and 56, there appear to be two sets of levels 
with quasi-spherical or weakly deformed structure,
associated with configurations $A$ and $B$.
All levels decrease in energy for 52--54, away from
closed shell, and rise again at 56 due to the
neutron $\nu(2d_{5/2})$ subshell closure. From 58, 
there is a pronounced drop in energy for the states of the 
$B$~configuration. At 60, the two configuration cross, 
indicating a Type~II QPT, and the ground state changes from 
$9/2^+_1$ to $5/2^+_1$, becoming the bandhead of a 
$K=5/2^+$ rotational band composed of  $5/2^+_1, 7/2^+_1, 
9/2^+_1, 11/2^+_1, 13/2^+_1$ states. The intruder 
$B$~configuration remains strongly deformed and the band 
structure persists beyond 60.
A similar trend is encountered in the
even-even Zr isotopes with the same
neutron numbers, as can be seen in Fig.~4.

\subsection{Intertwined QPTs in the Nb chain}

\hspace{12pt}
Intertwined QPTs are characterized by concurrent
Type~I and Type~II QPTs.
Fingerprints of Type~II QPT are evident in
the evolution of the normal-intruder
mixing across the Nb chain.
Fig.~10 shows the probability $b^2$ of the $\Psi_B$
component, Eq.~(\ref{Prob-ab-cm}), in the wave functions
of the ground state ($J^+_{gs}$) and first-excited
state ($7/2^+_1$), as a function of neutron number.
The rapid change in structure of $J^+_{gs}$ from the normal 
$A$~configuration at neutron number 52--58
(small $b^2$ probability),
to the intruder $B$~configuration
at neutron number 60--64
(large $b^2$) is clearly evident, signaling a Type~II QPT.
It should be noted that the ground state angular momentum
changes from $J^{+}_{gs}=9/2^{+}$
in $^{93-99}$Nb to $J^{+}_{gs}=5/2^{+}$ in $^{101-105}$Nb.
Such a possible change 
is a characteristic signature of 
Type~II QPTs in odd-mass nuclei,
unlike even-even nuclei
where the ground state
remains $0^+$ before and after the crossing.
The configuration change appears sooner in the
$7/2^+_1$ state, which changes to the $B$ configuration
already at neutron number 58.
Outside a narrow region near neutron number 60, where the 
crossing occurs, the two configurations are weakly mixed 
and the states retain a high level of purity.
A glance at Fig.~5(b) shows that
such a trend is similar to that
encountered for the $0^+_1$ and $2^+_1$  
states in the respective $_{40}$Zr cores.
\begin{figure*}
\centering
\includegraphics[width=1\linewidth]{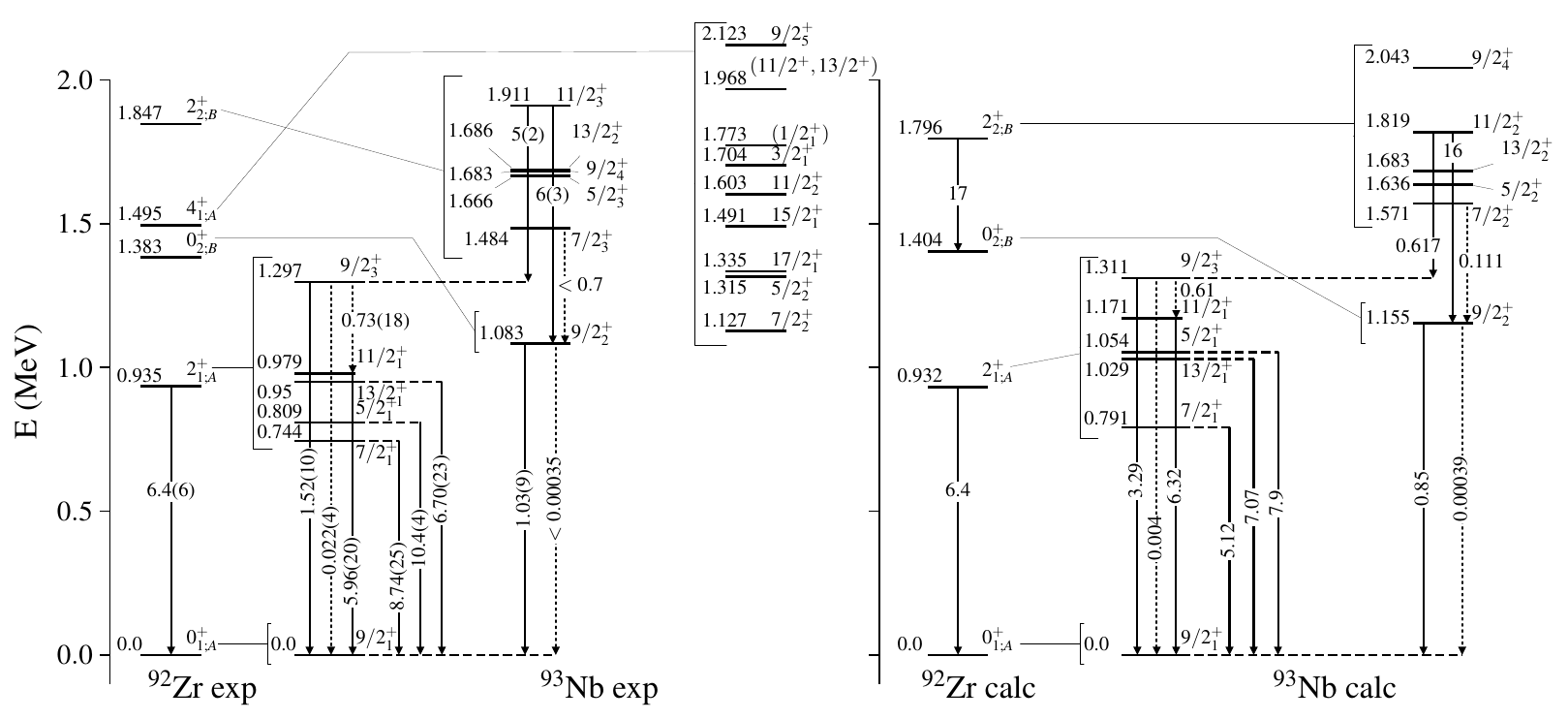}
\caption{
\small
  Experimental (left) and calculated (right) energy 
levels in MeV, and $E2$ (solid arrows) and $M1$ (dashed 
arrows) transition rates in W.u., for $^{93}$Nb and 
$^{92}$Zr. Lines connect $L$-levels in $^{92}$Zr to sets of 
$J$-levels in $^{93}$Nb, indicating the weak coupling 
$(L\otimes \tfrac{9}{2})J$, Eq.~(\ref{weak-coupl}).
Note that the observed $4^+_{\rm 1;A}$ 
state in $^{92}$Zr is outside the $N\!=\!1$ model space.
Adapted from~\cite{gavleviac22}.
\label{fig11:93Nb-p}}
\end{figure*}

\hspace{12pt}
Evidence for a \mbox{Type~I} QPT,
involving shape changes  
within the intruder $B$ configuration,
is obtained by examining the individual structure of
Nb isotopes at the end-points of the region considered.
Fig.~\ref{fig11:93Nb-p} displays the 
experimental and calculated levels in $^{93}$Nb along with 
$E2$ and $M1$ transitions among them. The corresponding 
spectra of $^{92}$Zr, the even-even core, are also shown  
with an assignment of each level $L$ to the normal $A$ or 
intruder $B$ configurations. The latter, as mentioned in
Section~2.5, are both spherical or weakly-deformed.
The low-lying states of the normal $A$ configuration
in $^{93}$Nb, 
can be interpreted in the weak coupling
scheme of Eq.~(\ref{weak-coupl}).
Specifically, the coupling of the single-proton
$\pi(1g_{9/2})$ state to the $L=0^+_{1;A}$ ground state
of $^{92}$Zr, yields the ground state $J=9/2^+_1$
of $^{93}$Nb. For 
$L=2^+_{1;A}$, it yields a quintuplet of states,
$5/2^+_1, 7/2^+_1, 9/2^+_3, 11/2^+_1,13/2^+_1$.
A multiplet built on  
$4^+_{1;A}$ can also be identified in the empirical 
spectrum of  $^{93}$Nb.
Particularly relevant to the present discussion is the
fact that the weak-coupling scheme is also valid for
non-yrast states of the intruder $B$ configuration in
$^{93}$Nb.
As shown in Fig.~\ref{fig11:93Nb-p},
the coupling of $\pi(1g_{9/2})$ to 
the $0^+_{2;B}$ state in $^{92}$Zr, yields the excited 
$9/2^+_2$ state in $^{93}$Nb and
for $2^+_{2;B}$~it yields the
quintuplet of states,
$5/2^+_3, 7/2^+_3, 9/2^+_4, 11/2^+_3, 13/2^+_2$.
\begin{figure}[t!]
\centering
\includegraphics[width=0.5\linewidth]{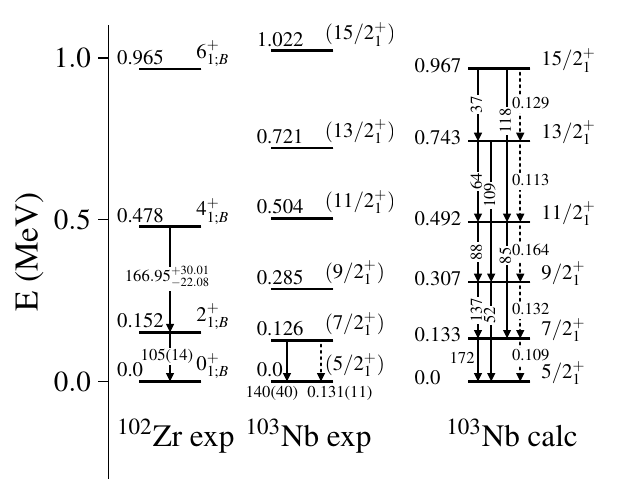}
\caption{
\small
  Experimental and calculated energy 
levels in MeV, and $E2$ (solid arrows) and $M1$ (dashed 
arrows) transition rates in W.u., for $^{103}$Nb and
$^{102}$Zr.
\label{fig12:103Nb-p}}
\end{figure}
\begin{figure*}[]
\begin{overpic}[width=0.49\linewidth]{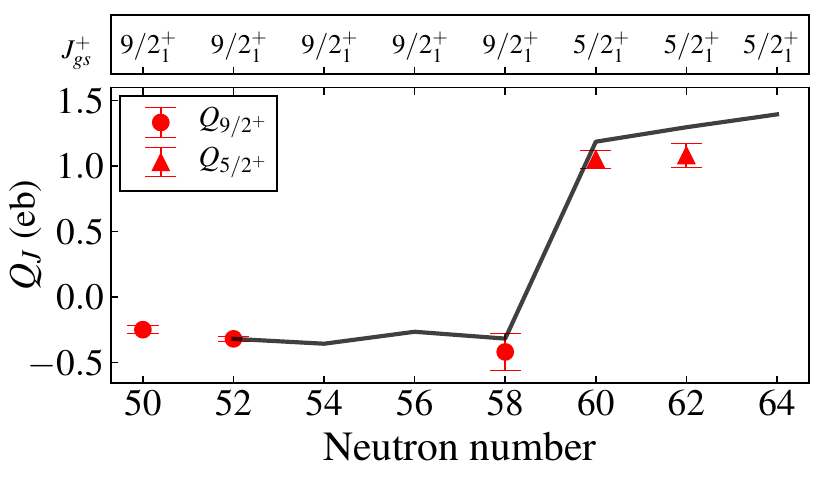}
\put (90,38) {(a)}
\end{overpic}
\begin{overpic}[width=0.49\linewidth]{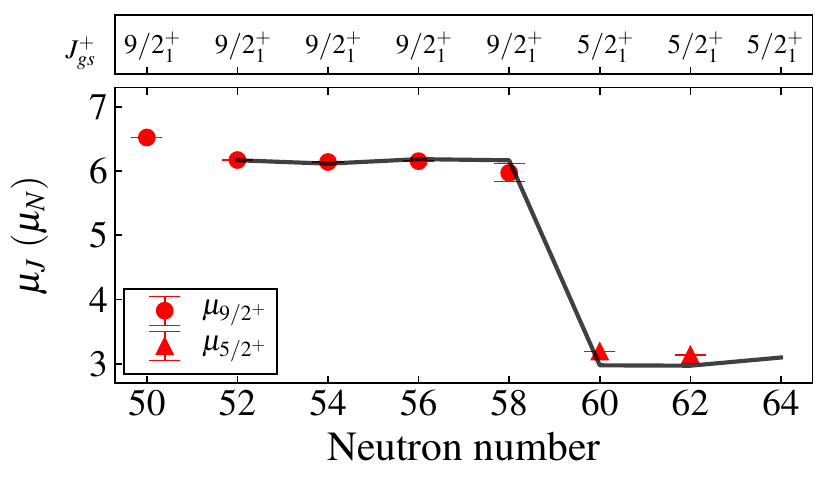}\\
\put (90,40) {(c)}
\end{overpic}\\
\begin{overpic}[width=0.49\linewidth]{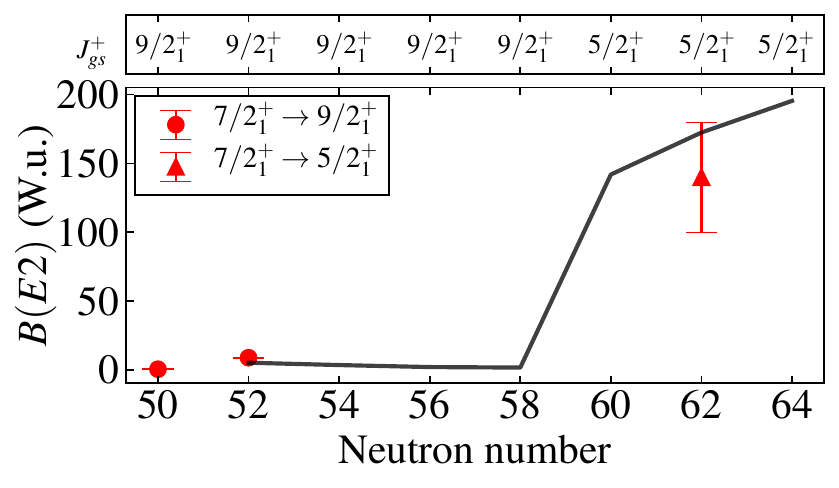}
\put (90,38) {(b)}
\end{overpic}
\begin{overpic}[width=0.48\linewidth]{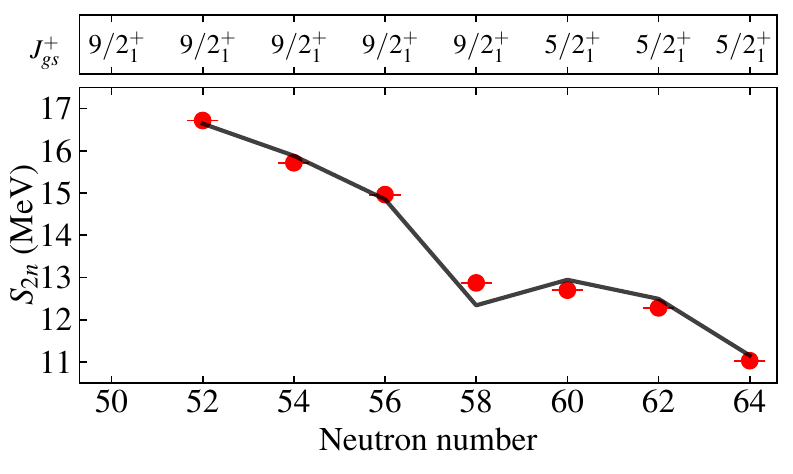}
\put (90,40) {(d)}
\end{overpic}
\caption{
\small  
Evolution of observables along the Nb chain.
Symbols (solid lines) denote experimental data 
(calculated results).
The values $J^+_{gs}$ for the ground state are indicated
at the top.
(a)~Quadrupole moments $Q_{J^+_{gs}}$ in $e$~barns ($eb$).
(b)~$B(E2; 7/2^+_1\!\to\! J^{+}_{gs})$ in W.u.
(c)~Magnetic moments $\mu_{J^{+}_{gs}}$ in $\mu_N$.
(d)~Two-neutron separation energy $S_{2n}$ in MeV.
}
\label{fig13:-order-p-be2}
\end{figure*}

\hspace{12pt}
For $^{103}$Nb, the yrast states, shown in
Fig.~\ref{fig12:103Nb-p},
are arranged in a $K=5/2^+$ rotational band
of a prolate-deformed core shape,
whose members
can be interpreted in the strong coupling scheme of
Eq.~(\ref{strong-coupl}).
The indicated states are obtained by 
coupling the proton $\pi(1g_{9/2})$ state to the
ground band ($L=0^+_{1;B},2^+_{1;B},4^+_{1;B},
6^+_{1;B},\ldots$) of the even-even core $^{102}$Zr,
also shown in Fig.~\ref{fig12:103Nb-p},
which is
associated with the intruder $B$ configuration. The 
calculations reproduce well the observed particle-rotor 
$J(J+1)$ splitting, as well as, the $E2$ and $M1$ 
transitions within the band. Altogether, we see an 
evolution of structure from weak-coupling of a spherical 
shape in $^{93}$Nb, to strong-coupling of a deformed shape 
in $^{103}$Nb. Such shape-changes within the 
$B$ configuration (Type~I QPT), superimposed on an abrupt 
configuration crossing (Type-II QPT), are the key defining 
feature of intertwined QPTs. Interestingly, this
intricate scenario, originally observed in the 
even-even Zr isotopes,
persists in the adjacent odd-even Nb~isotopes.
\subsection{Evolution of observables along the Nb chain}

\hspace{12pt}
Electromagnetic transitions and moments provide further 
insight into the nature of QPTs.
The quadrupole moment of the ground state $J^{+}_{gs}$
and $B(E2; 7/2^+_1\!\to\! J^{+}_{gs})$
in Nb isotopes are shown
in Fig.~13(a) and Fig.~13(b), respectively.
These observables are related to the deformation, the order 
parameter of the QPT. Although the data is incomplete, one 
can still observe small (large) values of these observables 
below (above) neutron number 60, indicating an increase in 
deformation. The calculation reproduces well this trend and 
attributes it to a Type~II QPT involving a jump between 
neutron number 58 and 60, from a weakly-deformed 
$A$ configuration, to a strongly-deformed
$B$ configuration. 
This behavior is correlated with 
a similar jump seen for the
B(E2)'s of $2^+\!\to\!0^+$ transitions in the even-even
Zr cores, shown in Fig.~6(a), and with the calculated
order parameters, shown in Fig.~5(a).
A jump in values at neutron number 60 is also seen in
the magnetic moment of the ground state ($\mu_{J^+_{gs}}$),
shown in Fig.~13(c). The evolution of the
two-neutron separation energy ($S_{2n}$) along the
Nb chain, shown in Fig.~13(d), exhibits features
of Type~I QPTs within the $A$ configuration
($B$ configuration) for neutron number 52-58 (60-64),
similar to the behaviour in the even-even Zr isotopes,
seen in Fig.~6(c).

\section{Concluding remarks}

\hspace{12pt}
We have presented the notion of intertwined
quantum phase transitions (QPTs) and considered its
manifestation in a system of bosons and in a mixed
system of bosons and fermions.
Intertwined QPTs correspond to
a situation of a QPT involving the crossing of two
(or more) configurations (Type~II), each of which is
in itself undergoing a QPT (Type~I).

\hspace{12pt}
As a case study of intertwined QPTs in a finite Bose
system, we have employed the algebraic
interacting boson model with configuration mixing,
describing collective states in even-even nuclei
in terms of bosons.
A comprehensive
analysis of spectral properties of Zr isotopes,
including the 
evolution of structure, order parameters and related
observables, revealed the role of two configurations.
The normal configuration remained spherical along
the Zr chain, while the intruder configuration
experienced first a transition from spherical [U(5)]
to an axially-deformed [SU(3)] shape, and then
a crossover to a $\gamma$-unstable [SO(6)] shape.
These gradual shape-phase transitions involve
a change of symmetries and are superimposed on an
abrupt crossing of the two configurations.
The normal-intruder mixing is weak (apart from the
crossing point), ensuring that both types of QPTs are
distinguishable, thus demonstrating intertwined QPTs
in even-even nuclei. 

\hspace{12pt}
As a case study of intertwined QPTs in a finite
Bose-Fermi system, we have employed the algebraic
interacting boson-fermion model with configuration
mixing, which describes low-lying states in odd-even
nuclei in terms of ground and core-excited bosonic
configurations coupled to a single-fermion.
An application to the Nb isotopes disclosed
a Type~II QPT (abrupt crossing of normal and intruder
states) accompanied by a Type~I QPT (gradual shape
evolution and transition from weak to strong coupling
within the intruder configuration), thus demonstrating
intertwined QPTs in odd-even nuclei.
The observed intertwined QPTs
in the odd-mass Nb isotopes echo the intertwined
QPTs of the adjacent even-even Zr isotopes.

\hspace{12pt}
The results reported in the present contribution
motivate further experiments on the spectroscopy  
of non-yrast states in even-even and odd-even nuclei,
as well as set the path for new investigations on
multiple and intertwined QPTs
in other Bose and Bose-Fermi systems.

\ack
A fruitful collaboration with N. Gavrielov (GANIL)
and F. Iachello (Yale) on the topics considered,
is acknowledged.

\end{document}